\documentclass[twocolumn,showpacs,preprintnumbers,footinbib]{revtex4}
\usepackage{bm,enumerate,dcolumn,tikz,graphicx,color,amsmath,amssymb}
\usepackage{epstopdf,empheq,xfrac,capt-of,pgfplots,array} 

\usepgfplotslibrary{colormaps,external}
\usetikzlibrary{decorations.markings,arrows} 
\usepackage{hyperref}

\newcommand{\comment}[1]{}
\newcommand{\bra}{\langle}
\newcommand{\ket}{\rangle}

\newcolumntype{C}[1]{>{\centering\let\newline\\\arraybackslash\hspace{0pt}}p{#1}}
\newcolumntype{L}[1]{>{\raggedright\let\newline\\\arraybackslash\hspace{0pt}}p{#1}}
\newcolumntype{R}[1]{>{\raggedleft\let\newline\\\arraybackslash\hspace{0pt}}p{#1}}

\makeatletter
\newenvironment{leftalign*}[1][\parindent]{\setlength\hangindent{#1}\start@align\tw@\st@rredtrue\m@ne}{\endalign}

\makeatother

\newcommand{\threej}[6]{
\begin{pmatrix}
  #1 & #2 & #3\\
  #4 & #5 & #6
\end{pmatrix}
}
\newcommand{\sixj}[6]{
\begin{Bmatrix}
  #1 & #2 & #3\\
  #4 & #5 & #6
\end{Bmatrix}
}

\begin{document}

\title{
  Tuning long-range interactions in Sr  Rydberg atoms: the effect of series perturbations 
} 

\author{Turker Topcu$^{1}$} 
\altaffiliation{Corresponding Author. Present address: Department of Mathematics, Virginia Tech, Blacksburg, Virginia 24061, USA }
\email[]{turkert@vt.edu}
\author{Andrei Derevianko$^{2}$} 
\affiliation{
  $^{1}$Department of Mathematics, Virginia Tech, Blacksburg, Virginia 24061, USA \\
  $^{2}$Department of Physics, University of Nevada, Reno, Nevada 89557, USA
}

\date{\today}

\begin{abstract}
We investigate the effect of series perturbation on the second-order  dipole-dipole interactions between strontium atoms in the $5sns({^1}S_0)$ and $5snp({^1}P_1)$ Rydberg states as a means of engineering long-range interactions between atoms. The series perturbation in these atoms enables modifying the strength and the sign of the interaction by varying the principal quantum number $n$ of the Rydberg electron. We utilize experimentally available data to estimate the importance of perturber states, and find that the van der Waals interaction between two strontium atoms in the $5snp({^1}P_1)$ states shows strong peaks that deviate from the anticipated hydrogenic $n^{11}$ scaling. We identify this to be the result of the perturbation of $5snd({^1}D_2)$ intermediate states by the $4d^2({^1}D_2)$ and $4dn's({^1}D_2)$ states in the $n<20$ range. This demonstrates that divalent atoms offer a unique advantage for generating substantially stronger or weaker inter-atomic interactions than those that can be achieved using alkali metal atoms. This is due to the highly perturbed spectra of divalent atoms and other multivalent atoms that can persist up to high $n$. Such irregularities can be especially useful in engineering asymmetric Ry blockade requiring the  simultaneous presence of both ``weak'' and ``strong'' interactions.

\vspace{0.5em}
\noindent {\bf Citation} T Topcu and A Derevianko 2026 J. Phys. B: At. Mol. Opt. Phys. {\bf 59} 215002 \\
\href{https://iopscience.iop.org/article/10.1088/1361-6455/aebb7c}{{\bf DOI}: 10.1088/1361-6455/aebb7c}
\end{abstract}


\maketitle

\section{Introduction}\label{sec:intro}
Long-range interactions between Rydberg (Ry) atoms are a useful resource in realizing conditional quantum dynamics, enabling a number of applications in quantum information processing (QIP) with neutral atoms~\cite{UrbJohHen09, GaeMirWil09, BetSafYak13, SafWalMol10}. Mediated by strong interactions between Ry atoms, the Rydberg blockade mechanism, which prohibits simultaneous excitation of two nearby Ry atoms, has particularly expanded the QIP toolbox. Applications initially focused on alkali-metal atoms, and have included  quantum logic gates~\cite{JakCirZol00, SafWalMol10}, simulation of exotic quantum many-body systems~\cite{MulLesWei09,WeiMulLes10}, study of strongly-correlated systems~\cite{PohDemLuk10,CinJaiBon10}, and multiparticle entanglement generation~\cite{SafMol09}. Strong Rydberg-Rydberg interactions in alkali metals  have also facilitated the realization of strongly interacting individual photons paving the way for quantum non-linear optics at the single photon level~\cite{PeyFirLia12, ChaVulLuk14}. 

Multivalent atoms, on the other hand, offer  new possibilities in engineering quantum systems~\cite{SafMol08, Olmos2013, Dunning2016, Kaubruegger2019, Chen2022}. Divalent atoms, such as group-II atoms ({\it e.g.}, Mg, Ca, Sr) and group-II-like atoms such as Yb, Hg, Cd, and Zn are the simplest examples of {\em multivalent} atoms. These  atoms possess the advantage of an extra valence electron, which makes them easier to trap  in tight optical lattices~\cite{MukMilNat11, Olmos2013, TopDer14, Yamamoto2016, Komar2016}, as well as in optical tweezer arrays~\cite{Wilson2022, Saskin2019}. This is greatly beneficial as optical trapping is essential for neutral-atom  QIP experiments due to the long coherence times that can be achieved. Coupled with mature experimental techniques for cooling and trapping divalent atoms~\cite{XuLofHal03, RapKriWas04, HemDraCha14, SorFerPol06}, divalent atoms offer additional advantages in QIP schemes compared to alkalis. Sub-microsecond qubit rotations and long coherence times in a universal set of quantum gate operations have already been experimentally demonstrated using Yb nuclear qubits in tweezer arrays~\cite{MaBuLi22, JeLiSe22} and Rydberg array based quantum simulators~\cite{ScShTs23}. 

These advantages have motivated extensive experimental and theoretical work on divalent Rydberg atoms, especially Sr and Yb. Recent spectroscopic studies have characterized Rydberg series in fermionic~\cite{DiWhKa18} and bosonic~\cite{CoNoHu19,Fields2021} Sr, in Yb~\cite{LeZuMa18}, and in open-shell erbium~\cite{TrMaIl21}. Rydberg blockade interactions have been investigated in ultracold Sr gases~\cite{DeSalvo2016,Qiao2021}, while collision-induced decoherence and loss have been studied in thermal Sr gases~\cite{Zhang2015,Yoshida2017,Hanley2017,Fields2019,Yoshida2020}, including experiments reaching very high principal quantum numbers, $n\sim 300$. In parallel, one- and two-qubit entangling gates with fidelities exceeding the surface-code threshold have been demonstrated with Yb~\cite{MaLiPe23,EvBlKa23}, and electromagnetically induced transparency measurements have shown long Rydberg coherence times~\cite{Gaul2016}.

Calculations of $C_6$ coefficients and polarizabilities based on multichannel quantum-defect theory for fermionic Sr~\cite{HuWeSe24,Rob19,RoBoSa18}  have shown that the hyperfine structure of the $^{87}$Sr$^+$ ionic core can strongly mix Rydberg series and produce nonmonotonic Rydberg-Rydberg interactions for $n>50$. This mechanism is absent in bosonic $^{88}$Sr due to its nuclear spin being zero. Similar nonmonotonic features have also been predicted for the rare-earth isotope $^{165}$Ho~\cite{RoBoSa18}. In bosonic Sr, calculations of van der Waals interactions between singlet $^{1}D_2$ and triplet $^{3}D_2$ $|5snd(|M_J|=2)\rangle$ Rydberg states have shown significant singlet-triplet mixing, leading to nonmonotonic $C_6$ coefficients for $n<25$ through singlet-singlet, singlet-triplet, and triplet-triplet intermediate pair states~\cite{Vaillant2015}. More broadly, interest in Rydberg-Rydberg interactions in alkaline-earth atoms has led to extensions of the open-source ARC package, originally developed for alkali-metal atoms, to include divalent Rydberg states~\cite{Robertson2021}. 

In this paper, we investigate long-range interactions between pairs of bosonic Sr atoms in the $|5sns(^{1}S_0)\rangle$ and $|5snp(^{1}P_1)\rangle$ Rydberg states. Because bosonic Sr has no hyperfine structure, it provides a useful setting in which to isolate the effects of series perturbations on van der Waals interactions. We also consider a one-dimensional optical-lattice geometry and derive $C_6$ coefficients for pairs of atoms either within the same lattice site or in different sites separated by a few lattice constants. The resulting expressions depend on an angle $\theta$ that specifies the orientation of the quantization axis relative to the internuclear axis: $\theta=0$ for atoms in different lattice sites and $\theta=\pi/2$ for atoms in the same lattice site (see Fig.~\ref{fig:1d_lattice}). We find pronounced nonmonotonic behavior in the $C_6$ coefficients for pairs of singlet $5snp(^{1}P_1)$ states even in the low-$n$, $n<20$, range. Working in this range may help mitigate collisional quenching of high-$n$ Rydberg states due to thermal motion~\cite{Fields2019, Hanley2017, Yoshida2017}. We further find that singlet-triplet mixing plays a relatively small role here compared with the $^{1}D_2$-$^{3}D_2$ interactions reported in~\cite{Vaillant2015}. Finally, we compare the van der Waals energy shifts with other relevant energy scales, including short-range interactions, quadrupole-quadrupole interactions, and Zeeman shifts, for realistic experimental parameters.

\begin{figure}[htb]
  \begin{center}
    \resizebox{0.7\columnwidth}{!}{\includegraphics{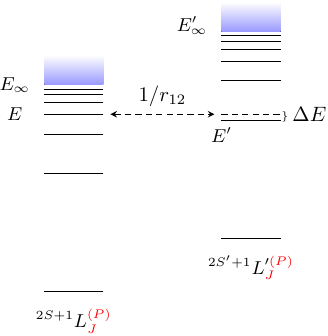}}
  \end{center}
  \caption{(Color online) Illustration of series interaction where two Rydberg series ${^{2S+1}}L^{(P)}_{J}$ and ${^{2S'+1}}L'^{(P)}_{J}$ with the same total $J$ and parity $P$ converge to different ionization thresholds $E_{\infty}$ and $E'_{\infty}$. If two states from these series happen to be close in energy (states labeled $E$ and $E'$), the Coulomb interaction $1/r_{12}$  can mix these states. The mixing is most prominent when the energy difference $\Delta E$ is small compared to the Coulomb interaction matrix element.
  }
  \label{fig:spert}
\end{figure}

The leading long-range interaction between Ry atoms in the same parity states is the second-order dipole-dipole (or the van der Waals) interaction. The van der Waals (vdW) interaction between alkali-metal atoms scales as $n^{11}$, $n$ being the principal quantum number of Ry electron. The multivalent atoms, however, can deviate significantly  from this scaling law, which is a feature unique to multivalent atoms, and  it is caused by the so-called Ry series interaction. The  highly-excited energy levels of divalent atoms can be characterized by the electronic configuration  $\gamma n \ell$, where $n$ and $\ell$  are principal and orbital quantum numbers of the Rydberg electron while, $\gamma$ denotes the set of quantum numbers  for the ``spectator'' electron. Consider two Ry series  $\gamma_1 n \ell$ and $\gamma_1' n' \ell'$ ($\gamma_1$ and $\gamma_1'$  are fixed, while  $n$ and $n'$  are scanned through the series for given  $\ell$ and $\ell'$).  Generically, such series exhibit the usual hydrogen-like behavior of energy levels. However, at some values of $n$ and $n'$, the two series may come close to being degenerate (see Fig.~\ref{fig:spert}). For such cases, if the symmetries (the total angular momenta and parities) of the two series are identical, neither of the states remains a ``good'' eigenstate of the atomic Hamiltonian and the levels are mixed by the off-diagonal Coulomb interaction between the configurations. This mechanism is well-known in spectroscopy~\cite{GallagherBook} and is referred to as the series interaction or series perturbation. 

As we demonstrate below, the series interaction leads to substantial deviations from the $n^{11}$ scaling law in Sr $|5snp(^{1}P_1)\rangle$ states with $n<20$ , with Ry states of the same series exhibiting both relatively small and large van der Waals interactions. The vdW interaction, which  scales with interatomic separation  as $R^{-6}$, arises in the second-order in the dipole-dipole interaction between two Ry atoms. As with  any second-order contribution, it is expressed as a sum over intermediate states with  energy denominators. The energy difference between the reference Ry+Ry states and an intermediate diatom state, entering the denominator, can deviate from the nominal behavior due to the series interaction of the intermediate state with another series of the same symmetry. Specifically, it is this perturbation of the intermediate states that causes the irregularities in the van der Waals interaction strengths we report here. Such irregularities can be especially useful in asymmetric Ry blockade~\cite{SafMol08} , which requires the  simultaneous presence of both ``weak'' and ``strong'' interactions.

 In this paper , we focus on Sr  due to  its common use in fermionic and bosonic  optical lattice clocks~\cite{AeKiWa24, ZhOoHi24, Akatsuka2008,DerKat11}, in addition to its potential for applications in Heisenberg-limited metrology through the use Rydberg blockade. Recently, entangled tweezer clocks utilizing partially entangled GHZ states have been  realized~\cite{FiTsSu24}. We show that the vdW interaction between two Sr atoms in the $5snp({^1}P_1)$ states display highly non-monotonic behavior for  low-$n$ states ($n\lesssim 20$), strongly deviating from the smooth, alkali-like $n^{11}$ scaling. We trace the origin of this behavior to unusually small energy denominators in the second-order energy expression resulting from the strong variation in atomic energies due to the series interaction. In particular, the $nP+nP\rightarrow nD+(n-1)D$ channel contributes to the $C_{6}$ coefficients with small energy denominators due to the highly perturbed nature of the $5snd({^1}D^{\rm e}_{2})$ series in Sr.  Based on spectroscopic data, we find that the $5snd({^1}D^{\rm e}_2)$ series is perturbed by the $4d^2({^1}D_2^{\rm e})$, $5p_{3/2}^2({^1}D_2^{\rm e})$ and $4d6s({^1}D_2^{\rm e})$ states, which lie below the ionization threshold of the $5snd({^1}D_2^{\rm e})$ series at 45,932 cm$^{-1}$~\cite{VaiJonPot14, Vaillant_2024}. Particularly, the $4d6s({^1}D_2^{\rm e})$ state perturbs states with $n=11-17$, $5p^2({^1}D_2^{\rm e})$ perturbs states with $n=5-6$ and $4d^2({^1}D_2^{\rm e})$ state perturbs the state with $n=12$ in the $5snd({^1}D_2^{\rm e})$ series. For the interaction between two Sr atoms in the $5sns({^1}S^{\rm e}_0)$ states, we do not see this non-monotonic behavior in the $n$-range we consider. Similar effects on dynamic polarizabilities in Rydberg electrons in fermionic Sr have been predicted  for some alkaline earth and lanthanide atoms for $n\geq 50$~\cite{Rob19, RoBoSa18}.

The paper is organized as follows: in Sec.~\ref{sec:spert}, we start with a brief description of the mechanism behind the series perturbation in divalent atoms and the one-dimensional lattice geometry we consider. Then in Sec.~\ref{sec:vdW}, we present our formalism and describe the computational framework. Results of our calculations  for  the ${^1}S_{0}+{^1}S_{0}$ and ${^1}P_{1}+{^1}P_{1}$ vdW interaction coefficients are presented in Sec.~\ref{sec:results}. Here we discuss the effects of the perturber states on the van der Waals interactions, which are pronounced for the $5snp({^1}P_1)$ Rydberg states with $n \lesssim 20$.  We also fit the scaled $C_6$ coefficients we calculated to rational functions of $n$. Fitting to rational functions with a polynomial of $n$ in the denominator has the advantage that they do not diverge as $n\rightarrow \infty$ and can potentially help mimic sharp nonmonotonic features in the data when fitted over a small $n$-range. We discuss the angular dependence of the vdW interactions in  Sec.~\ref{sec:angDist},  and compare the strengths  of the vdW  and quadrupole-quadrupole interactions in a typical one-dimensional lattice geometry in Sec.~\ref{sec:qq_interaction}. Finally, we  conclude  Sec.~\ref{sec:conc}. Atomic units ($\hbar=m_e=|e|\equiv 1$) are used throughout unless stated otherwise.

\section{Series perturbation in divalent atoms}\label{sec:spert}
We start by presenting a more detailed account of the series interaction mechanism and describe how it also results in van der Waals interactions that can strongly deviate from the $n^{11}$ alkali-metal-atom scaling law. Fig.~\ref{fig:spert} illustrates the basic qualitative picture. Two Rydberg series identified with the terms ${^{2S+1}}L^{(P)}_{J}$ and  ${^{2S'+1}}L'^{(P)}_{J}$ have identical total angular momenta $J$ and parities $P$. In the limit  $n \rightarrow \infty$  these series converge to different  thresholds  labeled as $E_{\infty}$ and $E'_{\infty}$. Two levels, $E$ and $E'$, of these series are nearly degenerate in energy, which can be quantified by comparing their energy separation $\Delta E$ to the off-diagonal matrix element of the Coulomb interaction evaluated between these nearly-degenerate states. Because of the accidental near degeneracy, the two levels repel each other, as long as  the Coulomb matrix elements do not vanish. The two series must have the same $J$ and $P$, otherwise the Coulomb matrix element would vanish. Typically, the Rydberg states lying above the perturbing state are shifted up in energy whereas the states below the perturbing state are shifted down~\cite{GallagherBook}.

For two identical atoms separated by a distance $R$ and in fixed magnetic sublevels, the vdW interaction reads
\begin{equation}\label{eq:vdW_noAngRed}
  \delta E_{\rm vdW}(R) = \sum_{j,k}
  \frac{| \bra\psi({\rm I})| \bra\psi({\rm II})| V_{DD}(R) |\phi_{j} ({\rm I})\ket |\phi_{k}(\rm{II})\ket |^2}{2E_{\psi} - (E_{j}+E_{k}) } \;.
\end{equation}
Here $\rm{I}$ and $\rm{II}$ label the two atoms.  The electronic  states of the two atoms, $\psi({\rm I})$ and $\psi({\rm II})$, are the same, and are either $|5sns({^1}S_{0})\ket$ or $|5snp({^1}P_{1})\ket$. The intermediate states $|\varphi_j\rangle$ and $|\varphi_k\rangle$ in Eq.~\eqref{eq:vdW_noAngRed} are unperturbed eigenstates of the atomic Hamiltonian. For the $|5sns\,^1S_0\rangle$ target states they run through the $|5sn'p\,^1P_1\rangle$ states, and for the $|5snp\,^1P_1\rangle$ target states through the $|5sn's\,^1S_0\rangle$ and $|5sn'd\,^1D_2\rangle$ states (here we assumed that the total spin is a good quantum number). The intermediate pair states are the products of these single-atom states, i.e., the $s-s$, $d-d$, and the $s-d$ pair channels, and each product enters the sum with its own energy denominator; the two possible assignments of the $s$ and $d$ states to the two atoms in the $s-d$ channel are accounted for by the explicit factor of 2 in the reduced sum $S_{sd}$, Eq.~\eqref{eq:pp_red_sum}. 

Because the LS-coupling  breaks down for some range of $n$ in Sr, some of these states have triplet character mixed into them as we will discuss below.  The dipole-dipole interaction between two atoms ($\rm I$ and $\rm II$) can be decomposed as
\begin{eqnarray}\label{eq:dd-pot}
  V_{DD}(R) &=& -\frac{1}{R^3} \sum_{\mu} w_{\mu}^{(1)}
  D_{\mu}^{(1)}({\rm I}) D_{-\mu}^{(1)}({\rm II}) \;,
\end{eqnarray}
where $w_{\mu}^{(1)}=1+\delta_{\mu,0}$, and the conventional electric-dipole operators are defined through their spherical-basis components as
\begin{eqnarray}\label{eq:dtensor}
  D_{\mu}^{(1)} &=& -\sum_{k} r_k C_{\mu}^{(1)}(\hat{\mathbf{r}}_k) \;.
\end{eqnarray}
Here $C_{\mu}^{(L)}(\hat{\mathbf{r}})=\sqrt{4\pi/(2L+1)}Y_{\mu}^{(L)}(\hat{\mathbf{r}})$ are the normalized spherical harmonics and $k$ runs over atomic electrons. In Eq. (\ref{eq:dd-pot}), the quantization axis is directed along the internuclear axis. Clearly, the vdW interaction~\eqref{eq:vdW_noAngRed} exhibits the $\propto R^{-6}$ scaling:
\begin{equation}\label{eq:vdW_r6}
  \delta E_{\rm vdW}(R) = \frac{1}{R^6} \sum_{j,k}
  \frac{| V_{j,k} |^2}{\Delta E_{j,k}} = -\frac{C_6}{R^6} \;.
\end{equation}
Here we replaced the numerator in~\eqref{eq:vdW_noAngRed} with $|V_{j,k}|^2/R^6$, and the denominator with $\Delta E_{j,k}$.

The expression~\eqref{eq:vdW_noAngRed} depends sensitively on the energy denominators, so accidental near-degeneracies between the target and intermediate states can result in strong deviations from the $n^{11}$ hydrogenic scaling of the long-range interaction strengths. As discussed below, we observe such an effect in the van der Waals interactions between two $5snp({^1}P_1)$ Sr Rydberg atoms,  as this interaction has an intermediate excitation channel $5snp({^1}P_1)+5snp({^1}P_1)\rightarrow 5sn'd({^1}D_2) + 5sn'd({^1}D_2)$ and the $5snd({^1}D_2)$ states are strongly perturbed below $n\simeq20$.

\begin{figure}[htb]
  \begin{center}
    \resizebox{0.9\columnwidth}{!}{\includegraphics{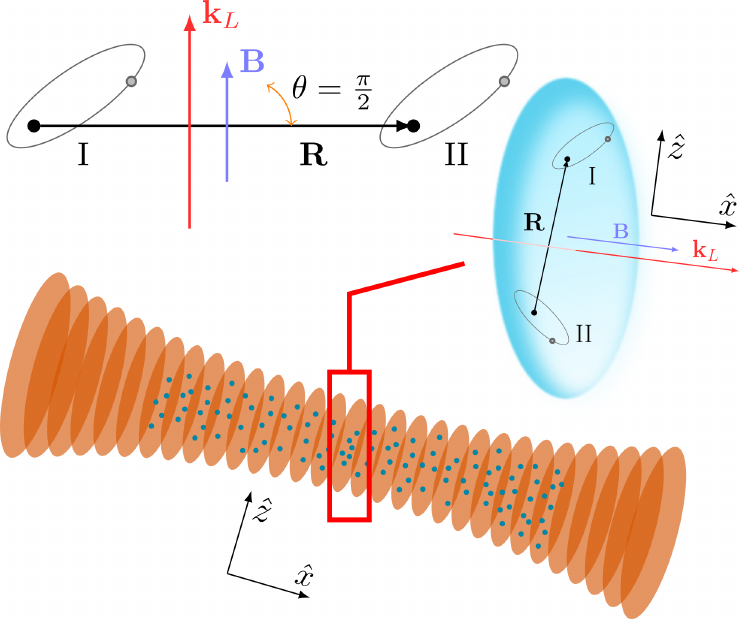}}
  \end{center}
  \caption{(Color online) One-dimensional lattice geometry described in the text in which atoms are trapped in  pancake-shaped  clouds at each lattice site. Application of a magnetic field $\mathbf{B}$ perpendicular to the plane of the pancake clouds lifts the Zeeman degeneracy where the  quantization  axis is perpendicular to the internuclear axis $\mathbf{\hat{R}}$ between atoms in each cloud. Here $\mathbf{k}_L$ is the wave vector of the lasers making up the optical lattice.
  }
  \label{fig:1d_lattice}
\end{figure}

In our calculations,  the LS-coupling scheme  does not hold  for all $n$-ranges in Sr, particularly in the  Ry series with high angular momentum $L$. For example, in the $5snd(D_2)$ Ry states, the series with the singlet and the triplet characters significantly mix for $n=15$ and 16, whereas the mixing is negligible for all other $n$ up to 30~\cite{VaiJonPot14}. In the case of the $5s15d(D_2)$ states, roughly 35\% of the channel is of the $5s15d({^1}D_2)$ character, and 55\% is of the $5s15d({^3}D_2)$ character. Other series of interest with lower angular momenta, {\it i.e.} $5sns(S_{0})$ and $5snp(P_{1})$, do not display this mixing in the $n$-ranges we are interested in this work , and we include the  mixing between the $5snd({^1}D_2)$ and $5snd({^3}D_2)$ series  in our calculations. Although  appreciable for $n=15$ and 16, this mixing  does not affect our qualitative conclusions as we will discuss in Sec.~\ref{sec:vdW}. We give details of how we include this effect in our calculations in Appendix~\ref{app:sing-trip}. 

To highlight the effect of series perturbation as clearly as possible, we evaluate the vdW coefficients $C_6$ for a nondegenerate manifold of magnetic substates. Zeeman degeneracy can be lifted by applying an external magnetic field, and we work with atoms trapped in a pancake-shaped cloud so that the interactions occur in the pancake plane (Fig.~\ref{fig:1d_lattice}). We assume a field strength for which the Zeeman splitting is large compared with the dipole-dipole couplings between different magnetic sublevels, allowing us to treat a single molecular channel. In particular, we focus on the stretched pair state with $\Omega_x = M_x + M_x^\prime = 2$, corresponding to $M_x=1$ and $M_x^\prime=1$, where $M_x$ and $M_x^\prime$ are the projections of the angular momenta of the two atoms along the magnetic-field direction. This choice isolates a single molecular channel and eliminates the need to treat mixing among different $M_x$ sublevels. In second-order perturbation theory, the relevant dipole-dipole couplings conserve the total projection $M_x+M_x^\prime$. As a result, the linear Zeeman shifts cancel in the energy denominators $\Delta E_{j,k}=2E_\psi-E_j-E_k$ and therefore do not affect the calculated $C_6$ coefficients. Thus, in the regime considered here, the magnetic field selects the quantization axis and suppresses sublevel mixing, but it does not otherwise enter the $C_6$ coefficients through the linear Zeeman shifts. 

For atom pairs within a single pancake-shaped cloud, the internuclear axis $\mathbf{\hat{R}}$ lies in the pancake plane, which we take to be the $y$-$z$ plane. Fig.~\ref{fig:1d_lattice} shows this trapping geometry, which is common in practical realizations of optical lattice clocks~\cite{BloNicWil14, HinShePhi13, LudBoyYe14}. An external magnetic field $\mathbf{B}$ can be applied along the lattice axis $\hat{x}$, making the quantization axis orthogonal to the internuclear axis $\mathbf{\hat{R}}$. For a sufficiently strong magnetic field, we may neglect vdW-induced mixing between magnetic sublevels and focus on a single $M_x$ sublevel, where $M_x$ is the projection of the atomic angular momentum along the magnetic-field direction. The quantization axis in Eq.~\eqref{eq:vdW_noAngRed}, however, is along the internuclear axis $\mathbf{\hat{R}}$. We therefore express the dipole-dipole interaction $V_{DD}$ in the magnetic-field frame by rotating the dipole operators $D_{\mu}^{(1)}$ in Eq.~\eqref{eq:dd-pot} through an angle $\theta$. This rotation is described by the Euler angles $\alpha=0$, $\beta=\theta$, and $\gamma=0$. If $\hat{S}$ denotes the corresponding rotation operator, the normalized spherical harmonics in Eq.~\eqref{eq:dtensor} transform as 
\begin{eqnarray}\label{eq:Ctensor}
  C_{M_K}^{(K)}(\hat{S}\hat{\mathbf{r}}) =
  \sum_{M=-K}^{K} C_{M}^{(K)}(\hat{\mathbf{r}})
  \mathbb{D}_{M,M_K}^{(K)}(0,\theta,0)  \;,
\end{eqnarray}
where $\mathbb{D}_{M,M'}^{(K)}(\alpha,\theta,\gamma)$ are the Wigner $D$-functions~\cite{VarMosKhe89}. This results in the rotated dipole operator
\begin{eqnarray} \label{eq:d1_rotated}
  D^{(1)}_{\mu}(\hat{S}\hat{\mathbf{r}}) =
  -\sum_{j} r_j
  \sum_{M=-1}^{1} C_{M}^{(1)}(\hat{\mathbf{r}}_j)
  \mathbb{D}_{M,\mu}^{(1)}(0,\theta,0) \;.
\end{eqnarray}
Here and below, $M_x$ denotes the projection of angular momentum along the lattice axis $\hat{x}$, which is also the magnetic-field direction for the $\theta=\pi/2$ geometry used in our same-site calculations. Unless otherwise stated, our numerical calculations use the pancake geometry shown in Fig.~\ref{fig:1d_lattice}. This is not the only geometry relevant to a one-dimensional lattice: for $\theta=0$, the two atoms occupy different pancake-shaped clouds, and the internuclear axis $\mathbf{\hat{R}}$ is parallel to $\hat{x}$. More generally, in two- and three-dimensional optical lattices, $\theta$ determines the relative orientation of the quantization and internuclear axes and therefore affects the van der Waals interaction between atoms in different lattice sites. We therefore keep the expressions in Appendix~\ref{app:vdw_derivation} explicit in $\theta$, so that they can be applied to geometries beyond the one shown in Fig.~\ref{fig:1d_lattice}. 

Finally, we discuss the magnetic field strength sufficiently large enough to suppress mixing between different magnetic sublevels, and restrict the calculation to a single $\Omega_x$ = 2 molecular channel. The required field strength follows from comparing the Zeeman splitting with the largest off-diagonal dipole-dipole coupling. At $\theta=\pi/2$, the rotated dipole-dipole operator couples the stretched $\Omega_x=2$ state to pair states with $\Omega_x=0$; near $n=18$ these couplings are of the same size as the near-resonant diagonal coupling, $V_{\mathrm{dd}}\approx 1.5$~GHz at the mean same-site separation $d\approx 320$~nm (Secs.~II.A and IV.B). The Zeeman splitting between the $\Omega_x=2$ and $\Omega_x=0$ manifolds is $\Delta E_Z/h=2(\mu_B/h)B\approx 2.8$~MHz/G$\times B$, so isolating the $\Omega_x=2$ channel near $n=18$ requires fields of order several hundred gauss. Away from the $n=18$ resonance, the off-diagonal couplings at $R=d$ are $\sim\!1$--$10$~MHz and $B\sim 1$--$10$~G suffices. At $\theta=0$ (atoms in different lattice sites), $V_{\mathrm{dd}}$ is exactly diagonal in $M_x$ and no field is required for channel isolation. 

\subsection{Length scales in one-dimensional lattice}\label{sec:length_scales}
In this subsection, we specify the relevant length scales of the one-dimensional optical lattice shown in Fig.~\ref{fig:1d_lattice}. These estimates are used below to compare competing interactions, including the quadrupole-quadrupole and van der Waals interactions. We use trap frequencies and atom numbers representative of current experimental setups. Establishing these length scales is useful for assessing quantum protocols that rely on long-range Rydberg interactions. 

Fig.~\ref{fig:1d_lattice} defines the trap geometry in a one-dimensional optical lattice. The lattice axis is along $\mathbf{k}_L$, and the in-plane internuclear axis between atoms within a pancake-shaped cloud is labeled $\mathbf{\hat{R}}$. The cloud volume, together with the number of atoms per lattice site, sets the mean interatomic separation within each site. To estimate this separation, we use typical trap frequencies $f_L$ and $f_R$ along $\mathbf{k}_L$ and $\mathbf{\hat{R}}$, respectively~\cite{BloNicWil14}, and assume approximately 20 atoms per lattice site. The mean separation places an upper bound on the accessible Rydberg principal quantum number: the Rydberg radius should remain less than roughly half the interatomic distance to avoid orbital overlap with neighboring atoms. 

For a harmonic trap, the oscillator length along direction $i$ is $a_i=\sqrt{\hbar/(m\omega_i)}$, with $\omega_i=2\pi f_i$. We use the corresponding full harmonic-oscillator widths $W_i=2a_i$ to estimate the volume of a pancake-shaped cloud. For $f_R=450$ Hz and $f_L=80$ kHz, this gives $W_R\simeq 1.0~\mu{\rm m}$ and $W_L\simeq 76$ nm~\cite{BloNicWil14}, with zero-point energy scales $\hbar\omega_R/(2k_B)\simeq 11$ nK and $\hbar\omega_L/(2k_B)\simeq 2~\mu{\rm K}$. For $N=10$ atoms per lattice site, we estimate the volume per atom as $V=\pi W_L(W_R/2)^2/N$, which gives a mean interatomic spacing $d\simeq(6V/\pi)^{1/3}\simeq 227$ nm. Requiring the Rydberg radius to remain below $d/2$ gives $n^*<32$. 

For a looser trap, this spacing increases; for example, $f_R=225$ Hz and $f_L=40$ kHz gives $d\simeq 320$ nm and permits $n^*<39$. Note that $N=10$ here refers to the number of Ry atoms per lattice site, and the actual number of atoms can be larger such that ground-state atoms lie inside Rydberg orbitals. In this case, short-range electron-atom interactions can form long-range Rydberg molecules and shift the Rydberg levels. For the parameters considered here, these shifts are typically at the MHz level~\cite{desalvo2015} and are small compared with the relevant energy denominators. At higher atom densities, multiple ground-state atoms can lie within a single Rydberg orbital, producing MHz-scale density-dependent broadening in addition to MHz-scale shifts~\cite{GajKruBal14}. 

A second relevant length scale is the lattice spacing $D=\lambda_{\rm m}/2$, which determines the separation between atoms in different lattice sites. For a Sr magic-wavelength lattice with $\lambda_{\rm m}=813$ nm, this gives $D\simeq 407$ nm.

\section{Van der Waals interactions}\label{sec:vdW}
Now we focus on computing the vdW interaction between two identical Rydberg atoms. Because it is second order in the dipole-dipole interaction, vdW contribution to the long-range molecular potential varies as $1/R^6$ with the inter-atomic separation and corresponds to the $-C_6/R^6$ term in the conventional multipole expansion of the long-range interactions. We are interested in the interactions between two $5sns({^1}S_{0})$ or two $5snp({^1}P_{1})$  atoms. When evaluating the interaction between the  $5snp({^1}P_{1,M_x})$ and $5snp({^1}P_{1,M'_x})$ atoms, we will consider the stretched  state $\Omega_x = M_x + M_x'=2$ where the only magnetic sublevels are $M_x=M_x'=1$. For two spherically-symmetric $5sns({^1}S_{0})$ atoms the rotation of quantization axis is irrelevant and $M_x=M_x'=0$. 

Derivation of the vdW interaction expressions in the LS coupling is given in Appendix~\ref{app:vdw_derivation} . The vdW interaction for two Sr atoms in the $5sns({^1}S_0)$ Rydberg states can be expressed as
\begin{align} \label{eq:c6-ss}
   & C_6
  \big (5sns({^1}S_0) (\text{I}) \, 5sns({^1}S_0)(\text{II}) \big ) = \nonumber \\
   & -\frac{2}{3}
  \sum_{\substack{{n_2,n_2'} }}
   \frac{
    \Big | \big\bra ns \big| \big| d \big| \big| n_2 p \big\ket \Big|^2 \;
    \Big | \big\bra ns \big| \big| d \big| \big| n_2' p \big\ket \Big|^2
  }{ 2E_{ns\, {^1}\!S_0}
    - \left (E_{n_2p\, {^1}\!P_1} + E_{n_2'p\, {^1}\!P_1} \right) }  \; ,
\end{align}
which involves the reduced matrix elements of the atomic dipole moment operators. The $n_2 p$ and $n'_2 p$ single-electron states reflect the electronic configurations of the intermediate dimer states  $|5sn_2p({^1}P_1)\ket_{\rm I} |5sn_2'p({^1}P_1)\ket_{\rm II}$.  Because we work in the LS coupling scheme, the contributions from the ${^3}P_1$ intermediate states are ignored because they contribute through spin-changing transitions, which are forbidden in the non-relativistic formalism. However, the breakdown of LS coupling and the significant singlet-triplet mixing affect some of the $5snd\,{}^1D_2$ intermediate states in the $5snp\,{}^1P_1+5snp\,{}^1P_1$ interaction, which we will discuss in detail in Sec.~\ref{subsec:sing-trip} and describe how we account for it in Appendix~\ref{app:sing-trip}. 

For the interaction between two $5snp({^1}P_1)$ atoms in the $M_x=M'_x=1$ magnetic substates,  \begin{eqnarray} \label{eq:c6-pp}
  & C_6
  \big (5snp({^1}P_{1,1_x}) (\text{I}) \, 5snp({^1}P_{1,1_x})(\text{II}) \big ) =  \nonumber \\
  & \frac{1}{36} S_{ss}
  +\frac{181}{3600} S_{dd}
  +\frac{11}{72} S_{sd} \;.
\end{eqnarray}
The structure of this expression reflects two possible single-atom dipole excitation channels $5snp \,{^1}\!P_1 \rightarrow 5sn_2s\,{^1}\!S_0$ and $5snp\,{^1}\!P_1 \rightarrow 5sn_2d \,{^1}\!D_2$ with the reduced sums $S_{ss}$, $S_{sd}$, and $S_{dd}$ defined in Eq.~\eqref{eq:pp_red_sum} in Appendix~\ref{app:sing-trip}.

\subsection{Uncertainty Estimates}\label{subsec:c6_err}
The van der Waals interaction energy in Eq.~\eqref{eq:vdW_r6} depends sensitively on the energy denominators, and experimental uncertainties in the measured energies can potentially affect our results. In our calculations of the $C_6$  coefficients  from Eq.~\eqref{eq:vdW_r6}, we take the uncertainty of the measured energy level data into account when available, and report the error bars resulting from the propagation of the energy uncertainties:
\begin{flalign}\label{eq:err_quadrature}
  &(\delta C_6)^2 = \sum_{j,k} \left( \left |\frac{2V_{j,k}}{\Delta E_{j,k}}\right | \delta(V_{j,k}) \right)^2 + \left( \left |\frac{V^2_{j,k}}{\Delta E_{j,k}^2}\right | \delta(\Delta E_{j,k}) \right)^2 \;, \\
  &\delta(\Delta E_{j,k}) = \sqrt{(2\delta(\Delta E_{\psi}))^2 + \delta(\Delta E_{j})^2 + \delta(\Delta E_{k})^2}
\end{flalign}
Here we add the uncertainties in quadrature assuming a normal distribution of measured values with standard deviations $\delta E_\psi$, $\delta E_j$, and $\delta E_k$. For example, when evaluating $\delta(\Delta E_{j,k})$ for the vdW interaction of two $5snp({^1}P_{1})$ states, $\delta(\Delta E_{\psi})$ is the energy uncertainty $\delta(\Delta E_{5snp({^1}P_{1})})$, and the uncertainties $\delta E_j$ and $\delta E_k$ are one of $\delta E_{5sns({^1}S_{0})}$ and $\delta E_{5snd({^1}D_{2})}$. Furthermore, we calculate the matrix elements using a one-electron model potential described in the next section; we will demonstrate that uncertainties in matrix elements in Eq.~\eqref{eq:err_quadrature}, which can be viewed as errors made in the matrix elements by using a model potential, can be ignored near energy degeneracies. Thereby,
\begin{flalign}\label{eq:err_quad_noVdd}
  &(\delta C_6)^2 \approx \sum_{j,k} \left( \left |\frac{V^2_{j,k}}{\Delta E_{j,k}^2}\right | \delta(\Delta E_{j,k}) \right)^2 \;.
\end{flalign}

In our calculations, the energies of the 5s$n$s(${^1}S_0$), 5s$n$p(${^1}P_1$) and 5s$n$d(${^1}D_2$) states in the range $6\le n\le 20$ come from Ref.~\cite{SanNav10}, which are data listed in the online NIST database. The data reported in~\cite{SanNav10} include  uncertainties , which we use to estimate the errors in our calculated $C_6$ coefficients in $6\le n\le 20$.
We use energies from~\cite{Esherick77} for 5s$n$s(${^1}S_0$) in the range $21\le n\le 40$, and from~\cite{BeiLucTim82} in $41\le n\le 70$. These older data do not come with experimental  uncertainties , which is why there are no error bars in Fig.~\ref{fig:C6_coeff_ss} for states with $n>20$.
For the 5s$n$p(${^1}P_1$) states, we use data from~\cite{Esherick77}, which include experimental energies for $21\le n\le 33$, and calculated energies for $34\le n\le 60$, which do not come with experimental uncertainities.
Finally, for the 5s$n$d(${^1}D_2$) states, we use experimental data from~\cite{Esherick77} and~\cite{BeiLucTim82} for the ranges $21\le n\le 51$ and $52\le n\le 70$ respectively. 
We prefer the energies of Ref.~\cite{Esherick77} in the lower-$n$ range because they cover all three series, $^1S_0$, $^1P_1$, and $^1D_2$, from the same set of measurements, so that correlated systematic errors in one direction partially cancel in the energy differences forming the denominators; the more precise data from Ref.~\cite{BeiLucTim82}, which does not include the ${}^1P_1$ series, is used in the higher-$n$ range. 

\subsection{The singlet-triplet mixing}\label{subsec:sing-trip}
Because LS coupling breaks down for $n\lesssim 20$ in the $5snd\,D_2$ series, our calculation of the $^{1}P_1+^{1}P_1$ interaction must include intermediate channels with mixed singlet and triplet character. This mixing noticeably modifies the scaled coefficients $\widetilde{C}_6=C_6/n^{11}$ (see Sec.~\ref{sec:results}) near $n=18$, although the resulting change does not alter the qualitative behavior of the interaction.  The dominant contribution arises from admixture of triplet character into nominal $^{1}D_2$ intermediate states~\cite{VaiJonPot14}. By contrast, admixture of singlet character into nominal triplet states gives only a small contribution, because the associated terms have larger energy denominators, as discussed below. Appendix~\ref{app:sing-trip} describes how this mixing is incorporated in the calculation. 

For the van der Waals interaction between two $5s18p$ $^{1}P_1$ atoms, the $^{3}D_2$ intermediate state lies 381.6 GHz below the corresponding $^{1}D_2$ state. As a result, channels containing one or two triplet $D_2$ intermediate states have larger energy denominators than the purely singlet channel: $\Delta E_{13}=2E_{{^1}P_1}-(E_{{^1}D_2}+E_{{^3}D_2})$ and $\Delta E_{33}=2E_{{^1}P_1}-(E_{{^3}D_2}+E_{{^3}D_2})$, compared with $\Delta E_{11}=2E_{{^1}P_1}-(E_{{^1}D_2}+E_{{^1}D_2})$. Although singlet-triplet mixing introduces additional nonrelativistic matrix-element contributions in the numerators, these contributions are suppressed by the larger denominators. The singlet-triplet and triplet-triplet channels therefore make negligible contributions to $\delta E_{\rm vdW}$ compared with the singlet-singlet channel. 

This suppression is evident in the dominant $n=18$ channel, $|n{^1}P_1,n{^1}P_1\rangle\rightarrow |n{^{1,3}}D_2,(n-1){^{1,3}}D_2\rangle$. The corresponding denominators are $\Delta E_{11}=5$, $\Delta E_{13}=377$, and $\Delta E_{33}=759$ GHz. Thus, replacing one or both singlet intermediate states by triplet states increases the denominator by 373 GHz or 754 GHz, respectively, strongly suppressing the mixed-channel contributions.

The situation is different for the van der Waals interaction between two $^{1}D_2$ states considered in~\cite{Vaillant2015}. In that case, the relevant comparison involves the channels $|n{^1}D_2,n{^1}D_2\rangle\rightarrow |n{^{1,3}}F_3,n{^{1,3}}F_3\rangle$. The singlet-singlet denominator is $\Delta E_{11}=2E_{{^1}D_2}-(E_{{^1}F_3}+E_{{^1}F_3})$, while the singlet-triplet and triplet-triplet denominators are $\Delta E_{13}=2E_{{^1}D_2}-(E_{{^1}F_3}+E_{{^3}F_3})$ and $\Delta E_{33}=2E_{{^1}D_2}-(E_{{^3}F_3}+E_{{^3}F_3})$. Because the $^{1}F_3$ and $^{3}F_3$ levels are close in energy, these three denominators remain comparable. For $n=16$, for example, $\Delta E_{11}=-8213$, $\Delta E_{13}=-8163$, and $\Delta E_{33}=-8113$ GHz, so replacing one or both singlet $F_3$ intermediate states by triplet states changes the denominator by only 50 or 100 GHz. These shifts are much smaller than the corresponding changes in the $^{1}P_1+^{1}P_1$ interaction, so the triplet-containing channels are not suppressed as strongly. 

The difference between the $^{1}P_1+^{1}P_1$ and $^{1}D_2+^{1}D_2$ interactions can be traced to smaller quantum defects in higher-angular-momentum Rydberg series. As a result, the relevant singlet and triplet states lie closer in energy, allowing the electron-electron interaction to couple these configurations more efficiently. For example, the energy splitting between the $5s16f$ $^{1}F_3$ and $^{3}F_3$ states is 50.4 GHz, about seven times smaller than the 381.6 GHz separation between the $^{1}D_2$ and $^{3}D_2$ intermediate states relevant to our $n=18$ calculation. Consequently, the singlet-singlet, singlet-triplet, and triplet-triplet denominators in the $n\,^{1}D_2+n\,^{1}D_2$ interaction remain similar in magnitude, so mixed-channel contributions are not suppressed by large changes in the denominators. This explains why singlet-triplet mixing gives sizable contributions in~\cite{Vaillant2015}, whereas it plays a more muted role in the present $^{1}P_1+^{1}P_1$ results.

\subsection{Model Potential}\label{subsec:mod-pot}
While computing the one-electron reduced matrix elements of the dipole operator, we use the three-parameter model potential of Ref.~\cite{Millen11} describing the  interaction of the Ry electron with the residual atomic core,
\begin{equation}\label{eq:modelPot}
  U_{l}(r) = Be^{-Cr} - \frac{1 + (Z-1)e^{-Ar}}{r} + \frac{l(l+1)}{2r^2} \,.
\end{equation}
In Ref.~\cite{Millen11} the parameters ($A$, $B$, $C$) of this model potential were obtained by fitting its eigenspectrum to experimental data~\cite{Esherick77, BeiLucTim82} to minimize the differences between numerical and experimental energies for various terms such as ${^1}S_0$, ${^1}P_1$, ${^1}D_2$.  Because we need accurate energy denominators in \eqref{eq:c6-ss} and \eqref{eq:pp_red_sum}, we fit the energy denominators from the model potential to the experimental energy denominators, rather than fitting the energies from the model potential to experimental energies. The potential quoted in Ref.~\cite{Millen11} yields differences between the numerical and experimental energies that are comparable to the experimental energy denominators. Therefore we have modified some of the parameters in~\cite{Millen11} to match the experimental energy denominators at high-$n$ ($n\gtrsim 30$); the revised set of parameters is listed in Table~\ref{table:eff_potential}. The parameters marked with asterisks are those that are modified from the ones tabulated in Ref.~\cite{Millen11}. 

For example, in the range $20\le n\le 50$, the original parameters from Ref.~\cite{Millen11} for the ${^1}S_0$ series lead to fractional differences between the experimental and the calculated energies, $\delta E/E^{\rm exp}_{ns}$, at the $10^{-4}$ level for the 5s$ns$ series, and $10^{-2}$ level for the 5s$np$ series. Here $\delta E = E^{\rm calc} - E^{\rm exp}$ is the difference between the energy of a level computed with the model potential and its experimental value, both measured from the ionization limit. Our modified parameters seen in  Table~I, not only improve $\delta E/E^{\rm exp}_{ns}$ for the 5s$np$ series to $10^{-3}$, but also improve the error $\Delta\mathcal{E}^2(nS;\, n_2\ell_2,\,n_2'\ell_2')$ in the energy denominators in the channel $nS+nS\rightarrow nP+(n-1)P$ to $10^{-10}$ a.u.$^2$ from $10^{-7}$ a.u.$^2$ in the same range $20\le n\le 50$, where  
\begin{align*}
\Delta\mathcal{E}^2(nS;\,& n P,\,(n-1) P) \\
 & = \sum_n\left(\Delta E_{n P,\,(n-1) P}^{\exp }-\Delta E_{n P,\,(n-1) P}^{\text {calc }}\right)^2 \;, 
\end{align*}  
and $\Delta E_{n P,\,(n-1) P}$ is the energy denominator in the channel $nS+nS\rightarrow nP+(n-1)P$. The quantities $\Delta E^2(nP; n^2\ell^2, n'^2\ell'^2)$ are defined analogously to the equation above; for example, 
\begin{align*}
\Delta \mathcal{E}^2(nP;\, & nD, (n-1)D) \\
& = \sum_n (\Delta E^{\rm exp}_{nD,(n-1)D} - \Delta E^{\rm calc}_{nD,(n-1)D})^2 \;, 
\end{align*}  
where $\Delta E_{nD,(n-1)D} = 2E_{nP} - E_{nD} - E_{(n-1)D}$ is the energy denominator in the channel $nP+nP \rightarrow nD+(n-1)D$ in the ${}^1P_1+{}^1P_1$ interaction. We see a significant improvement in the $\Delta\mathcal{E}^2(nP;\, n_2\ell_2,\,n_2'\ell_2')$ error in energy denominators generated from the ${^1}P_1$ parameters in Table~\ref{table:eff_potential}, from $10^{-7}$ a.u.$^2$ to somewhere ranging between $10^{-9}$ and $10^{-10}$ a.u.$^2$ depending on the channel. In the dominant channel $nP+nP\rightarrow nD+(n-1)D$, we attain $\Delta\mathcal{E}^2(nP;\, nD,\,(n-1)D)\sim 10^{-9}$ a.u.$^2$ for $20\le n\le 50$. The fractional differences between the experimental and the calculated energies, $\delta E/E^{\rm exp}_{np}$ and $\delta E/E^{\rm exp}_{nd}$, remain at the same level $\sim\!10^{-2}$ as those calculated using the parameters from~\cite{Millen11}. 

The effects of perturbers on the 5s$n$s(${^1}S_0$), 5s$n$p(${^1}P_1$) and 5s$n$d(${^1}D_2$) series are analyzed based on a Multichannel Quantum Defect Theory (MQDT) approach in Ref.~\cite{VaiJonPot14} which reports channel fractions in these series. For example, in the 5s$n$s(${^1}S_0$) series, the main perturbing channels are 4d$n$d(${^1}S_0$) and 4d$n$d(${^3}P^e_0$), which account for much less than a percent in the character of the 5s$n$s(${^1}S_0$) states for $n>8$. In the 5s$n$p(${^1}P_1$) series, 4d$n$p(${^1}P_1$) accounts for less than 5\% for $n>10$. Finally, in the 5s$n$d(${^1}D_2$) series, 4d$n$s(${^1}D_2$) have channel fractions below 4\% for $n>10$, in addition to 5p$n$p which accounts for much less than a percent. Mixing from the triplet character through the 4d$n$s(${^3}D_2$) is also below one percent except for $n=15$ and 16 after which its channel fraction drops to well below a percent. Of particular  interest  to us in our Fig. 4 is $n=18$: mixing from the 4d$18$s(${^1}D_2$) is 2\% whereas all the other perturbers contribute less than one percent.

To take the perturbed nature of the Ry series into account, we evaluate the $C_{\rm 6}$ coefficients using experimental energies in the denominator of Eq.~\eqref{eq:c6-ss} and~\eqref{eq:c6-pp}. However, the energy spectrum of the model potential~\eqref{eq:modelPot} is smooth, and it cannot reproduce the perturbed energy levels. Therefore it cannot reflect the effect of the perturbers on the radial matrix elements. The model potential~\eqref{eq:modelPot} is similar to the one in Ref.~\cite{VaiJonPot12, Millen11} where the authors compare calculations of $C_6$ coefficients using the Coulomb approximation~\cite{WeTrMe17} and a model potential like the one we use in our calculations. They conclude that the radial matrix elements obtained using this model potential and the Coulomb approximation typically differ by about $\sim$0.1 a.u., and that this translates to about 0.5\% difference between the $C_6$ coefficients for $n\sim$20.

Furthermore, comparison with experimental Stark maps of Ry states has shown that treating a Ry state of an alkaline-earth metal in the single-electron model to calculate the dipole matrix elements~\cite{ZhChCh01} yields satisfactory results for $n<20$, specifically down to $n=12$ for Sr. We therefore expect the single-active-electron approach to be adequate for calculating the dipole matrix elements in this work. 

The radial wave functions in the Coulomb approximation are obtained for hydrogen using the correct experimental energies and integrating the radial Shr\"odinger equation~\cite{ZhChCh01}. This gives the correct energy and the large-$r$ behavior, but incorrect behavior near the core. For the dipole matrix elements, the correct energy and the asymptotic behavior are important. In case the error introduced by the Coulomb approximation in Ref.~\cite{VaiJonPot12} dominates the total error, in Appendix~\ref{app:matel_err}, we estimate how large the errors in our calculated matrix elements need to be to increase the sizes of the error bars to effectively recover the $n^{11}$ monotonic scaling.

In Appendix~\ref{app:matel_err}, we  express the error in our dipole matrix elements as an uncertainty $\delta(\nu_n)$ in the quantum defects $\nu_n=n-n^*$. We then show that $\delta(\nu_n)$ must be significantly larger than the variation of the quantum defects across the $^1P_1$ and $^1D_2$ Ry series reported in~\cite{VaiJonPot12} if the error bars are to become large enough to include the monotonic behavior. Furthermore, $\delta(\nu_n)$ must be also significantly larger than the difference between the calculated quantum defects using the model potential~\eqref{eq:modelPot} and the quantum defects from experimental data. 

\begin{center}
  \begin{figure}[h!tb]
    \captionof{table}{Parameters $A$,$B$, and $C$ of the model potential~\eqref{eq:modelPot} for the Rydberg electron in the $5sns(^1S_0)$, $5snp(^1P_1)$ and $5snd(^1D_2)$ states  of strontium. Most of the parameters are from Ref.~\cite{Millen11}, except the ones marked with asterisks, which were determined in this work by minimizing the differences between the experimental and the numerical values of energy denominators in the second order $^1\!S_0 + ^1\!S_0$ (top) and $^1\!P_{1} + ^1\!P_{1}$ (bottom) van der Waals interactions. \\}
    \begin{tabular}{|c||c|c|c|}
      \hline
       & \multicolumn{3}{c|}{$5sns(^1S_0) + 5sns(^1S_0)$}             \\
      \cline{2-4}
       & $A$                                              & $B$ & $C$ \\
      \hline\hline
      $^1S_0$
       & 3.762                                                        
       & -6.33
       & 1.07                                                         \\ 
      \hline
      $^1P_1$
       & 2.84$^*$
       & -1.86
       & 1.10                                                         \\ 
      \hline
      $^1D_2$
       & 2.78                                                         
       & -9.06
       & 2.31                                                         \\ 
      \hline 
    \end{tabular}
    \begin{tabular}{|c||c|c|c|}
      \hline
       & \multicolumn{3}{c|}{$5snp(^1P_1) + 5snp(^1P_1)$}             \\
      \cline{2-4}
       & $A$                                              & $B$ & $C$ \\
      \hline\hline
      $^1S_0$
       & 5.599$^*$                                                    
       & -6.33
       & 1.07                                                         \\ 
      \hline
      $^1P_1$
       & 3.49
       & -1.86
       & 1.10                                                         \\ 
      \hline
      $^1D_2$
       & 2.588$^*$                                                    
       & -9.06
       & 2.31                                                         \\ 
      \hline
    \end{tabular}
    \label{table:eff_potential}
  \end{figure}
\end{center}

\section{Results}\label{sec:results}
Because the numerator of $\delta E_{\rm vdW}$ is proportional to the fourth power of the dipole operator ($\propto n^8$), while the energy denominator scales as $1/n^3$ for large $n$, the $C_6$ coefficient is expected to scale as $n^{11}$. This scaling is exact for hydrogen, but nonhydrogenic atoms exhibit residual deviations because the Rydberg electron interacts with a polarizable, penetrable ionic core. We therefore report scaled coefficients, $\widetilde{C}_6=C_6/n^{11}$. Any remaining $n$-dependence in the scaled coefficient reflects atomic and state-dependent structure beyond the hydrogenic scaling. Two types of residual behavior are relevant: a smooth monotonic trend, associated with core polarization, core penetration, and exchange effects, and resonance-like structure, which occurs when near-degenerate intermediate channels produce anomalously small energy denominators and enhanced interactions. The latter nonmonotonic features, especially in the low-$n$ region, are the focus of this paper. 

\begin{figure}[h!tb]
  \begin{center}
    \resizebox{1.0\columnwidth}{!}{\includegraphics{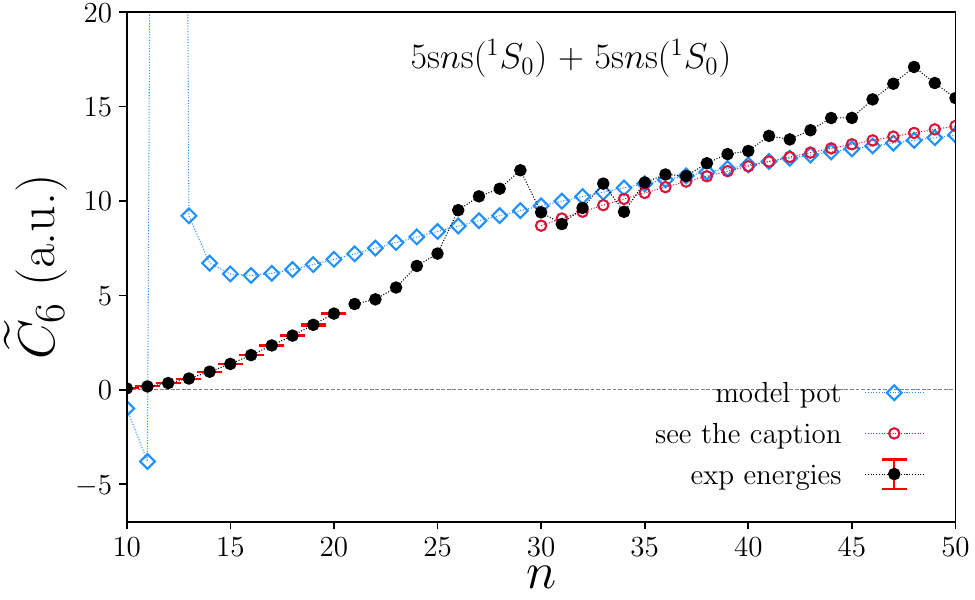}}
  \end{center}
  \caption{(Color online) Residual $n$-dependence of the $C_6$ coefficients ($\tilde{C}_6 = C_6/n^{11}$) for the $5sns({^1}\!S_0) + 5sns({^1}\!S_0)$ van der Waals interaction in Sr. The blue open diamonds are calculated using the model potential quoted in the text. These match the  theoretical calculations  from Ref.~\cite{VaiJonPot12} obtained using experimental energies  (red open circles) in the high-$n$ region. Replacing numerical energies from the model potential in the energy denominator with experimental energies results in solid black points. Positive values of $\widetilde{C}_6$ coefficients imply attractive vdW interactions. Error bars are calculated for $n<20$, beyond which no experimental  energy uncertainties  are reported.
  }
  \label{fig:C6_coeff_ss}
\end{figure}

\begin{figure}[h!tb]
  \begin{center}
    \resizebox{1.0\columnwidth}{!}{\includegraphics{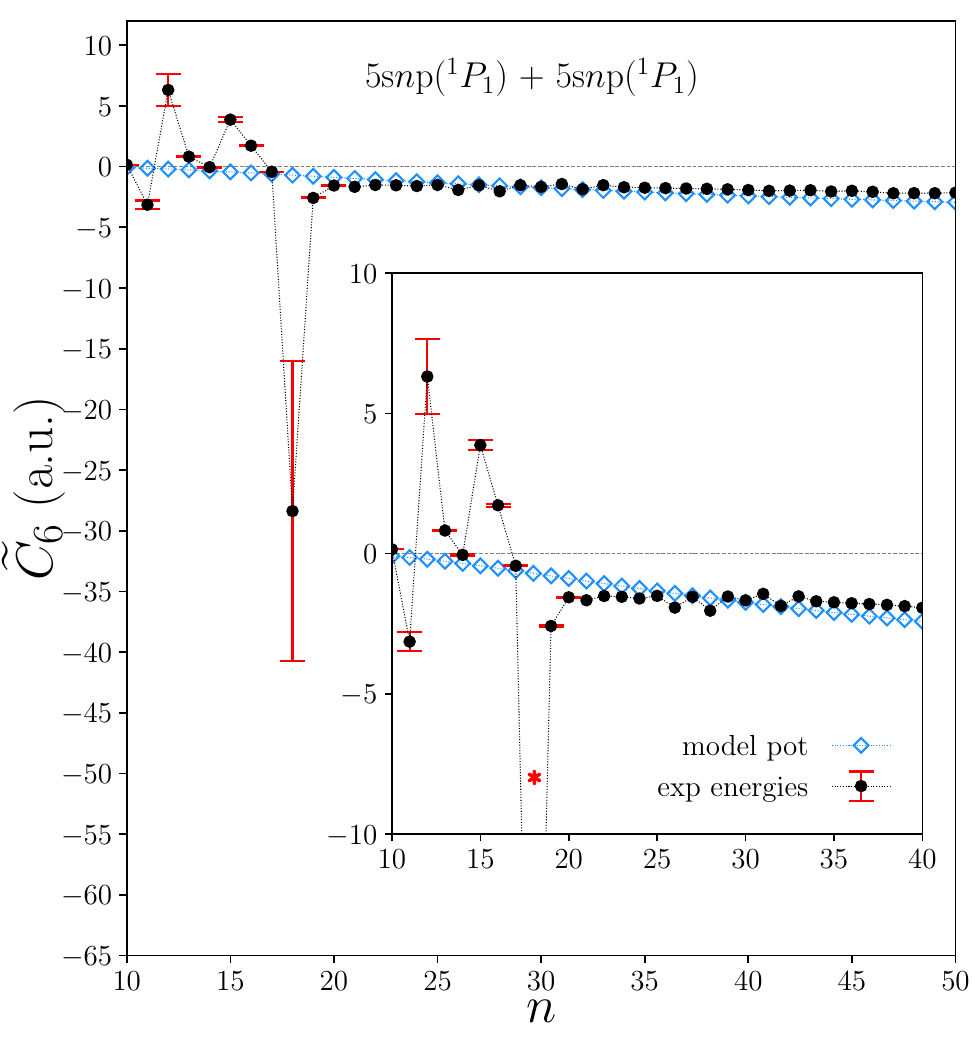}}
  \end{center}
  \caption{(Color online) Similar to Fig.~\ref{fig:C6_coeff_ss} but for the $5snp({^1}\!P_{1,1_x}) + 5snp({^1}\!P_{1,1_x})$ interaction. The blue open diamonds are evaluated using matrix elements and energies obtained from the model potential and solid black points using experimental energies in the energy denominator. The red star denotes the data point for $n=18$ outside the plot range: $\widetilde{C}_6 (n=18)\simeq -28$ a.u. The large error bar at $n=18$, like the large deviation in the $\widetilde{C}_6$ value, stems from near degeneracy in the energy denominator. 
  }
  \label{fig:C6_coeff_pp}
\end{figure}

The scaled $\widetilde{C}_6=C_6/n^{11}$ coefficients for the van der Waals interaction between pairs of $5sns({^1}S_0)$ and $5snp({^1}P_{1,1_x})$ atoms are plotted as a function of $n$ in Figs.~\ref{fig:C6_coeff_ss} and~\ref{fig:C6_coeff_pp}. We choose to scale by $n^{11}$ rather than $(n^*)^{11}$ because this gives us a smaller slope for the residual monotonic $n$-dependence in Figs.~\ref{fig:C6_coeff_ss} and~\ref{fig:C6_coeff_pp} since $n>n^*$.

\begin{table}[h]
  \centering
  \begin{tabular}{c c r r}
    \hline \hline
    \multicolumn{1}{c}{$n$}                       &
    \multicolumn{1}{c}{$\widetilde{C}_6$}         &
    \multicolumn{1}{r}{ $\delta\widetilde{C}_6$ } &
    \multicolumn{1}{r}{$|\delta\widetilde{C}_6 / \widetilde{C}_6|$}          \\
    \hline                                                                          \\[-3ex]
    10                                                   &  0.15  &  0.0003  &  0.002  \\
    12                                                   &  6.32    &  1.33    & 0.21   \\
    14                                                   &  -0.044   &  0.037   &  0.84    \\
    16                                                   &  1.72    &  0.058    &  0.03    \\
    18                                                   &  -28.38  &  12.38   & 0.44   \\
    20                                                   &  -1.56   &  0.015   & 0.01   \\
    \hline\hline
  \end{tabular}
  \caption{\label{Table:errBars} Tabulated values of the $\widetilde{C}_6$ coefficients for the $5snp({^1}\!P_{1,1_x}) + 5snp({^1}\!P_{1,1_x})$ interaction (Fig.~\ref{fig:C6_coeff_pp}), the propagated uncertainties from the energy denominators $\delta\widetilde{C}_6$, and the  corresponding  fractional uncertainties $|\delta\widetilde{C}_6 / \widetilde{C}_6|$ in the range of $n$ for which there is  uncertainty  data in~\cite{SanNav10}. }
\end{table}

\subsection{$C_6$ for the $5sns({^1}S_0)$ states}
In Fig.~\ref{fig:C6_coeff_ss}, open blue diamonds show the $\widetilde{C}_6$ coefficients calculated using Eq.~\eqref{eq:c6-ss} in which both the energy denominator and the one-electron orbitals in the reduced matrix elements are calculated using the model potential~\eqref{eq:modelPot}. The parameters in this model potential are adjusted from those listed in Ref.~\cite{Millen11} to better  match the experimental energy denominators. This curve matches the data reported in Ref.~\cite{VaiJonPot12} well (open red circles) for all $n$ quoted in~\cite{VaiJonPot12}. However, around $n\approx 12$, the $\widetilde{C}_6$ coefficients become extremely large due to an almost vanishing energy denominator. Keeping in mind that the energies used in generating this set of $\widetilde{C}_6$ were calculated from the model potential~\eqref{eq:modelPot}, we find that the fitted potential has the pathological behavior at these low-$n$ such that the energy denominator flips sign going from $n=12$ to 11. This results in an unphysical peak at $n=12$. A side effect of this is the overestimation of the $\widetilde{C}_6$ coefficients for $n$ up to around $25$ due to the large contribution from the unphysical $n=12$ intermediate state to the $\widetilde{C}_6$ coefficients of the nearby states.

On the other hand, replacing the numerically calculated energies in the denominator of~\eqref{eq:c6-ss} by experimental values~\cite{Esherick77, BeiLucTim82, SanNav10} results in the solid black points. While matching the other two sets of data well at high $n$, these keep monotonically decreasing for $n$ below 20. We display error bars indicating the propagated uncertainties in the calculated $\widetilde{C}_6$ coefficients in the range $n\le 20$ where reported experimental data in Ref.~\cite{SanNav10} also includes the uncertainties. Although these error estimates do not account for errors in the matrix elements incurred using a model potential as discussed in~\ref{subsec:mod-pot}, they show that the uncertainties which stem from the energy denominators are very small.

\subsection{$C_6$ for  the $5snp({^1}P_{1,1_x})$ states}
Unlike the $5sns({^1}S_0) + 5sns({^1}S_0)$ interaction, numerically calculated values of $\widetilde{C}_6$ using the model potential~\eqref{eq:modelPot} between $5snp({^1}P_{1,1_x})$ states do not display the same pathological behavior at low $n$ (open blue diamonds in Fig.~\ref{fig:C6_coeff_pp}). Replacing the numerically calculated energies in the denominators of Eq.~\eqref{eq:c6-pp} with experimental ones listed in~\cite{Esherick77, BeiLucTim82, SanNav10} changes the $\widetilde{C}_6$ coefficients very little for $n\gtrsim 20$ (solid black points).

\begin{figure}[h!tb]
  \begin{center}
    \resizebox{1.0\columnwidth}{!}{\includegraphics{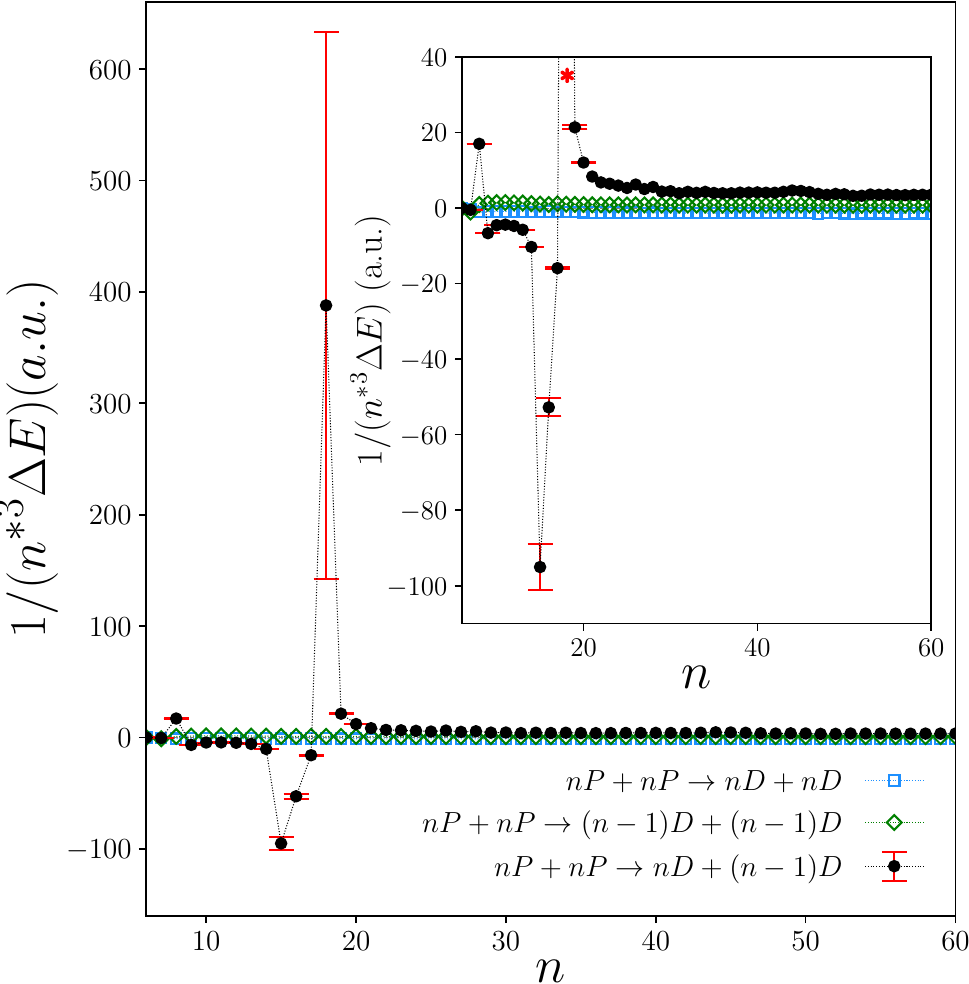}}
  \end{center}
  \caption{(Color online) The scaled energy denominator in the second order perturbation theory expression for the $\widetilde{C}_6$ coefficients in the  long-range  interaction between two $5snp({^1}P_{1,1_x})$ Sr atoms. $\Delta E$ scales as $1/{n^{*}}^{3}$ as expected for the $nP+nP\rightarrow (n-1)D+(n-1)D$ and $nP+nP\rightarrow nD+nD$ interaction channels over the entire $n$-range. For the $nP+nP\rightarrow nD+(n-1)D$ channel, however, this scaling breaks down for $n\lesssim 20$ and $1/({n^{*}}^{3} \Delta E)$ attains large values and changes sign. This ``resonance-like" structure is facilitated by an unusually small $\Delta E$ around $n\sim 17$ and is possible because of the highly perturbed nature of the $5snd({^1}D_{2})$ series in Sr. The error bars for the $nP+nP\rightarrow nD+(n-1)D$ channel show that taking the experimental  uncertainties  into account does not change the qualitative picture for $n<20$.
  }
  \label{fig:EnrDominator}
\end{figure}

We observe an unusual feature in the low-$n$ region below $n=20$: the $\widetilde{C}_6$ coefficients display a non-monotonic behavior. 
Unlike the unphysical peak at $n=12$ in the ${^1}S_0 + {^1}S_0$ interaction (Fig.~\ref{fig:C6_coeff_ss}), which stems from the inaccurate representation of the level energies by the model potential~\eqref{eq:modelPot} at low $n$, the non-monotonic features in the ${^1}P_{1,1_x} + {^1}P_{1,1_x}$ are real: they are etched in energy denominators derived from experimental spectra. 
For example, in the interaction between two $5s18p({^1}P_{1,1_x})$ states, the energy denominator in Eq.\eqref{eq:c6-pp} involving the $5s18p({^1}P_{1,1_x})$, $5s18d({^1}D_{2})$ and $5s17d({^1}D_{2})$ states becomes unusually small, which gives the large peak at $n=18$. The behavior of the energy denominator in the $nP+nP\rightarrow n'D+n''D$ channel can be seen in Fig.~\ref{fig:EnrDominator}.  To  emphasize its non-monotonic nature, we have plotted the scaled energy denominator $1/({n^{*}}^{3}\Delta E)$ rather than $\Delta E$. Here $\Delta E$ is the energy difference $2E_{nP}-(E_{n'D}+E_{n''D})$, and $n^{*}=n-\nu_n$ is the effective principal quantum number where the quantum defect $\nu_n$ is calculated using the Rydberg-Ritz formula
\begin{equation}\label{eq:ry-ritz}
  \nu_n = \nu^{(0)}_{n} + \frac{\nu^{(2)}_{n}}{(n-\nu^{(0)}_{n})^2} + \frac{\nu^{(4)}_{n}}{(n-\nu^{(0)}_{n})^4} \;.
\end{equation}
Here the coefficients $\nu^{(0)}_n$, $\nu^{(2)}_n$, and $\nu^{(4)}_n$ are obtained from fitting to experimental data in the range $10\le n\le 29$ for the $5snp({^1}P_{1})$ series, and $20\le n\le 50$ for the $5snd({^1}D_{2})$ series (Table 2 in~\cite{VaiJonPot12}).

Notice that the energy denominators associated with the channels $nP+nP\rightarrow (n-1)D+(n-1)D$ and $nP+nP\rightarrow nD+nD$ are  well-behaved  for all $n$ (open green diamonds and open blue squares). This is to be expected from a hydrogen-like spectrum where the nearest-neighbor energy spacing scales as $1/n^3$ for all $n$. On the other hand, whereas $1/({n^{*}}^{3}\Delta E)$ is flat above $n\sim 20$, it is far from regular below $n\sim 20$ in the $nP+nP\rightarrow nD+(n-1)D$ channel. Particularly at $n=18$, the energy difference $2E_{nP}-(E_{nD}+E_{(n-1)D})$ is unusually small resulting in the large peak in $\widetilde{C}_6$. Besides being flat above $n\sim 20$, $1/({n^{*}}^{3}\Delta E)$ is also much smaller than the values at the peaks below $n\sim 20$, which translates into large $\widetilde{C}_6$ coefficients for the van der Waals interactions between the ${^1}P_{1,1_x}$ states in the region $n\lesssim 20$. The error bars in Figs.~\ref{fig:C6_coeff_pp} and~\ref{fig:EnrDominator} show that taking the experimental  uncertainties  into account does not change the qualitative picture for $n<20$.  It is also worth reiterating that the open data points for $\widetilde{C}_6$ in Fig.~\ref{fig:C6_coeff_pp} were  obtained using experimental energies but do not include configuration interaction for the matrix elements.

The deviation of the nearest-neighbor energy differences from the hydrogenic $1/{n^*}^3$ scaling is a consequence of the series perturbation. The energies of the $5sns({^1}S_0)$ states almost perfectly scale as $1/{n^{*}}^2$ for all ${n^*}$, whereas $5snp({^1}P_{1})$ and $5snd({^1}D_2)$ states increasingly deviate from this scaling as ${n^*}$ gets below $\sim$20. This is because the states with higher angular momenta are more likely to be perturbed by other series. For a Rydberg series to be perturbed by another state, the energy of the perturber state needs to be below the ionization threshold for the Rydberg series. Furthermore, it needs to have the same $J$ and the parity for the associated Coulomb integral to be non-zero. Researching the literature, we have found that the $5snd({^1}D_2)$ series of Sr is perturbed by the $4d^2({^1}D_2^{\rm e})$, $5p_{3/2}^2({^1}D_2^{\rm e})$ and $4d6s({^1}D_2^{\rm e})$ states, all of which lie below the ionization threshold of the $5snd({^1}D_2^{\rm e})$ series at 45,932 cm$^{-1}$~\cite{VaiJonPot14}. Other candidates with higher angular momenta that could perturb the $5snd({^1}D_2^{\rm e})$ series are $4dng({^1}D_2^{\rm e})$; however, we could not find  experimental  data to confirm that these states lie below 45,932 cm$^{-1}$. We find that the $4d6s({^1}D_2^{\rm e})$ state perturbs states with $n=11-17$, $5p^2({^1}D_2^{\rm e})$ perturbs states with $n=5-6$ and $4d^2({^1}D_2^{\rm e})$ state perturbs the state with $n=12$ in the $5snd({^1}D_2^{\rm e})$ series. A detailed analysis of the $5sns({^1}S_0)$, $5snp({^1}P_1)$ and $5snd({^1}D_2)$ series and their perturbers is given in~\cite{VaiJonPot14}. In contrast with the $5snp({^1}P_{1,1_x}) + 5snp(^1P_{1,1_x})$ interaction, we see no contribution from perturber states for the $5sns({^1}S_0) + 5sns({^1}S_0)$ interaction at low-$n$ due to lower angular momentum of the perturber states in the ${^1}S_0$ series~\cite{VaiJonPot14}.

Furthermore,  the sum in Eq.~\eqref{eq:vdW_r6} is essentially dominated by one term with the smallest energy denominator. The energy denominator of this term determines the sign of the $C_6$ coefficient. The numerator is positive and, therefore, does not change the sign of the dominant term. The result is that the sign changes seen in the $C_6$  coefficients  in Fig.~\ref{fig:C6_coeff_pp} cannot be undone by changing the values of the dipole matrix elements. Therefore, the nonmonotonic behavior seen in Fig.~\ref{fig:C6_coeff_pp} stems from the energy denominator and would not disappear if more accurate radial matrix elements were used in our calculations. 

Because the small energy denominators strongly shape the $\widetilde{C}_6$ coefficients in Fig.~\ref{fig:C6_coeff_pp}, one might worry that degenerate perturbation theory is required. However, even the smallest denominators encountered here remain large enough that nondegenerate perturbation theory is well justified. The smallest denominator in Fig.~\ref{fig:EnrDominator} occurs for a pair of $5s18p\, {^1}\!P_1$ states in the $n<20$ region. Specifically, the $5s18p\, {^1}\!P_1+5s18p\, {^1}\!P_1\rightarrow 5s18d\, {^1}\!D_2+5s17d\, {^1}\!D_2$ channel gives the largest contribution, with an experimental energy denominator of about 5 GHz~\cite{SanNav10}. The relevant expansion parameter is the ratio of the dipole-dipole coupling to this energy separation,
\[
\eta = 
\frac{\bigl|\langle 5s18p\,{}^1\!P_1,\,5s18p\,{}^1\!P_1
|V_{\rm dd}|
5s18d\,{}^1\!D_2,\,5s17d\,{}^1\!D_2\rangle\bigr|}{R^3\,|\Delta E|} \ll 1,
\]
where $\Delta E$ is the energy difference between the initial and intermediate pair states. This expansion parameter $\eta = C_{3}/(R^{3}\Delta E)$ depends on the interatomic separation $R$, and therefore, on the trap configuration. Using either the computed two-electron dipole matrix element $C_{3}\approx 4.9\times 10^{4}$~a.u., or $C_{3}\approx 3.6\times 10^{4}$~a.u. inferred from Table~\ref{Table:errBars} with $\Delta E\approx 5$~GHz, we obtain $\eta\approx 0.3$ at the mean atomic separation $d\approx 320$~nm ($10$ atoms per site), and $\eta\sim 0.1$ at the inter-site spacing in a magic wavelength lattice $D\approx 407$~nm (Sec.~\ref{sec:length_scales}). Whether nondegenerate perturbation theory is quantitatively sufficient thus depends on the number of atoms per site and the trap parameters: at $d\approx 320$~nm and $D\approx 407$~nm the $R^{-6}$ description is appropriate, while at shorter separations of $d\approx 227$~nm, {\it e.g.} due to a tighter trap configuration with $f_{r}=450$~Hz, $f_{L}=80$~Hz, $\eta\approx 0.9$ is marginal. The $C_{6}$ coefficient itself is well-defined as the leading term of the long-range expansion; the nonmonotonic $n$-dependence seen in Fig.~\ref{fig:C6_coeff_pp} reflects genuine atomic structure (the perturbed $5snd$ series).

\subsection{Fitting $\widetilde{C}_6$ to rational functions of $n$}
For use in applications, we fit the scaled $\widetilde{C}_6$ coefficients to rational functions of $n$. Polynomial fits of second or third degree are often used for such data~\cite{VaiJonPot12}, but they can behave poorly outside the fitted range and diverge as $n\rightarrow \infty$. We therefore fit the model-potential results, shown as blue diamonds in Fig.~\ref{fig:C6_coeff_pp}, to rational functions that approach a finite limit. Specifically, we use $(an+b)/(n+d)$ and $(an^2+bn+c)/(n^2+en+f)$, where $a$, $b$, $c$, $d$, $e$, and $f$ are fitting parameters. The parameters are listed in Table~\ref{table:Cn_fits}, and both forms approach $a$ in the high-$n$ limit. A rational function can also mimic localized nonmonotonic behavior when fitted over a restricted $n$ interval; however, the fits reported here use only $n>20$ data and therefore do not reproduce the low-$n$ nonmonotonic feature below $n\sim 20$. 

Results of our least-squares fit for the  $\widetilde{C}_6$  coefficients are tabulated in Table~\ref{table:Cn_fits}. The parameters in Table~\ref{table:Cn_fits} indicate that, in the limit $n\rightarrow \infty$, the  van der Waals interaction between pairs of $5sns({^1}S_{0})$ atoms is roughly a factor of 3 stronger than that between pairs of $5snp({^1}P_{1,1_x})$ atoms. On the other hand, it is clear from Fig.~\ref{fig:C6_coeff_pp} that picking $n=18$ results in much stronger van der Waals interactions between the $5snp({^1}P_{1,1_x})$ states than what would be expected from the fits quoted in Table~\ref{table:Cn_fits}.

\begin{center}
  \begin{figure}[h!tb]
    \captionof{table}{Fit parameters for the scaled $\widetilde{C}_6$ for
    fits of the form $(an + b)/(n + d)$ and $(an^2 + bn + c)/(n^2 + en + f)$ for the ${^1}S_{0}+{^1}S_{0}$ and ${^1}P_{1}+{^1}P_{1}$ van der Waals interactions. The $\chi^2$ parameter for the lower order fit is $\lesssim$0.05 and $\chi^2 < 10^{-4}$ in the higher order fit. One needs to go to $n=200$ for the difference between the two fits for the $\widetilde{C}_6$  coefficients to become $\sim$10\%. 
    \\}
    \begin{tabular}{|c||c|c|c|}
      \hline
       & \multicolumn{3}{c|}{$(an + b)/(n + d)$}             \\
      \cline{2-4}
       & $a$                                     & $b$ & $d$ \\
      \hline\hline
      $\widetilde{C}_6({^1}S_0 + {^1}S_0)$   
       & 27.268                                              
       & -116.556                                            
       & 41.926                                              \\ 
      \hline
      $\widetilde{C}_6({^1}P_{1,1_{x}} + {^1}P_{1,1_{x}})$   
       & -11.036                                             
       & 112.5                                               
       & 97.176                                              \\ 
      \hline
    \end{tabular} \\
    \begin{tabular}{|c||c|c|c|c|c|}
      \hline
       & \multicolumn{5}{c|}{$(an^2 + bn + c)/(n^2 + en + f)$}                         \\
      \cline{2-6}
       & $a$                                                   & $b$ & $c$ & $e$ & $f$ \\
      \hline\hline
      $\widetilde{C}_6({^1}S_0 + {^1}S_0)$    
       & 22.821                                                                        
       & -473.021                                                                      
       & 3136.018                                                                      
       & 6.791                                                                         
       & -129.797                                                                      \\ 
      \hline
      $\widetilde{C}_6({^1}P_{1,1_{x}} + {^1}P_{1,1_{x}})$       
       & -6.342                                                                        
       & 97.465                                                                        
       & -390.021                                                                      
       & 22.228                                                                        
       & 260.072                                                                       \\ 
      \hline
    \end{tabular}
    \label{table:Cn_fits}
  \end{figure}
\end{center}

\section{Angular dependence of the van der Waals interaction}\label{sec:angDist}
 A detailed  derivation of the angular dependence of van der Waals interaction can be found in Appendix~\ref{app:vdw_derivation}. The dependence of the $\widetilde{C}_6$ coefficients for the ${^1}P_{1,1_x}+{^1}P_{1,1_x}$ van der Waals interaction on the angle $\theta$ that the quantizing  B-field makes with the internuclear axis can be expressed in terms of  Legendre polynomials $P_{0}(\cos\theta)$, $P_{2}(\cos\theta)$ and $P_{4}(\cos\theta)$. The angular part of the interaction energy $\delta E_{\rm vdW}$~\eqref{eq:denr} is contained in the square-bracketed terms in Eq.~\eqref{eq:vdW_gen} and can be written as
\begin{subequations}
  \label{eq:f_ang}
  \begin{eqnarray}
    f_{ss} (\theta) &=&
    \frac{1}{45} + \frac{2}{63} P_{2}(\cos\theta) + \frac{2}{35} P_{4}(\cos\theta)
    \label{eq:f_ss} \,, \\
    f _{dd}(\theta) &=&
    \frac{271}{4500} + \frac{32}{1575} P_{2}(\cos\theta) + \frac{1}{1750} P_{4}(\cos\theta)
    \label{eq:f_dd} \,,\\
    f_{sd} (\theta) &=&
    \frac{23}{225} -  \frac{61}{630} P_{2}(\cos\theta) + \frac{1}{175} P_{4}(\cos\theta)
    \label{eq:f_sd} \;,
  \end{eqnarray}
\end{subequations}
for the three channels involved, where we have denoted the angular part of the vdW interaction energy in Eq.~\eqref{eq:vdW_gen} with the functions $f$. Finally, we express the $C_6$ coefficients from~\eqref{eq:vdW_gen} using $\delta E_{\rm vdW}=-C_6/R^6$:
\begin{equation}\label{eq:c6_angFunc}
  C_6  =  f_{ss} (\theta) S_{ss} + f_{dd} (\theta) S_{dd} + f_{sd} (\theta) S_{sd} \;.
\end{equation}
Here the reduced sums $S_{ss}$, $S_{dd}$ and $S_{sd}$ are defined in Eq.~\eqref{eq:pp_red_sum}.

\begin{figure}[h!tb]
  \begin{center}
    \resizebox{0.9\columnwidth}{!}{\includegraphics{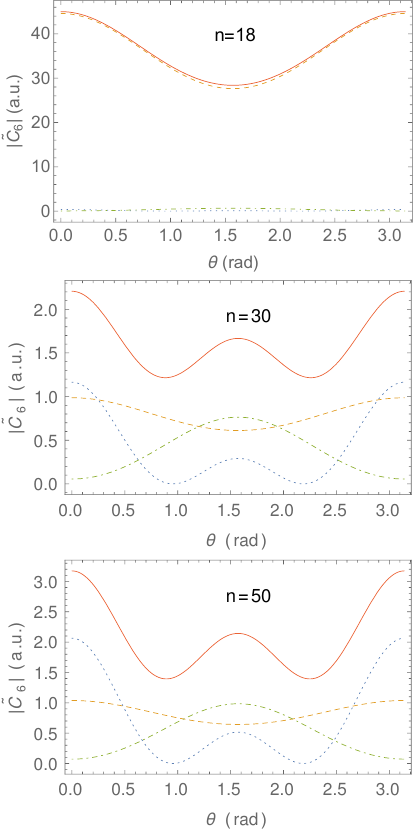}}
  \end{center}
  \caption{(Color online) Angular distributions of the $|\widetilde{C}_6|$ coefficients for the van der Waals interaction between two $5snp({^1}P_{1,1_x})$  Sr atoms (solid  orange  curves) for various $n$, and individual contributions from different intermediate channels:
  $5sns({^1}S_0) + 5sn's({^1}S_0)$ (blue dotted),
  $5snd({^1}D_2) + 5sn'd({^1}D_2)$ (dashed  orange), and
  $5sns({^1}S_0) + 5sn'd({^1}D_2)$ (dotted dash  green).
  Note that the $5sns({^1}S_0) + 5sn's({^1}S_0)$ channel is dominated by $P_{4}(\cos\theta)$ whereas the $5snd({^1}D_2) + 5sn'd({^1}D_2)$ channel is dominated by $P_{2}(\cos\theta)$. It is clear from the top panel that $\widetilde{C}_6$ is entirely determined by the $P_{2}(\cos\theta)$ character of the $5snd({^1}D_2) + 5s(n-1)d({^1}D_2)$ channel at $n=18$, where the $5snd({^1}D_2)$ series is strongly perturbed.
  }
  \label{fig:C6_angDist}
\end{figure}

The contributions from these three channels to the angular distribution of the coefficients $|\widetilde{C}_6|$ for three $n$ values are plotted in Fig.~\ref{fig:C6_angDist}. The $nP + nP\rightarrow n'S + n''S$ channel contribution is depicted by the blue dotted curves, the $nP + nP\rightarrow n'D + n''D$ channel by the  orange  dashed and the $nP + nP\rightarrow n'S + n''D$ channel by the  green  dot-dashed curves. The total is plotted as the solid  orange  curve in each case. It is clear from Eqs.~\eqref{eq:f_ang} and Fig.~\ref{fig:C6_angDist} that the $nP + nP\rightarrow n'S + n''S$ channel is dominated by $P_{4}(\cos\theta)$ whereas the $nP + nP\rightarrow n'D + n''D$ channel is dominated by $P_{2}(\cos\theta)$. All three channels contribute comparably to the total $\widetilde{C}_6$ over the range of all angles for $n=30$ and 50 which are states in the unperturbed part of  the Rydberg series with $\widetilde{C}_6$ behaving monotonically in Fig.~\ref{fig:C6_coeff_pp}. For $n=18$ however, the total $\widetilde{C}_6$ is entirely determined by the $nP + nP\rightarrow n'D + n''D$ channel. This is because the resonance-like peak at $n=18$ in Fig.~\ref{fig:C6_coeff_pp} is due to the highly perturbed nature of the $5snd({^1}D_2)$ Rydberg series which  results  in an unusually small energy denominator for $n=18$ in the $nP + nP\rightarrow n'D + n''D$ channel. The results we quote in Fig.~\ref{fig:C6_coeff_pp} and in Table~\ref{table:Cn_fits} are for $\theta=\pi/2$ which corresponds to the middle peak in the solid  orange  curves in Fig.~\ref{fig:C6_angDist}.

\section{Quadrupole-Quadrupole interactions}\label{sec:qq_interaction}
In neutral-atom quantum computing, long-range interactions are mediated by multipole couplings, and the dominant term can be either the van der Waals or quadrupole-quadrupole interaction, depending on the relevant length scale. Appendix~\ref{sec:qq_derivation} gives the derivation of the quadrupole-quadrupole interaction used below. 

We now compare the quadrupole-quadrupole interaction $E_{QQ}$ with the van der Waals interaction $E_{\rm vdW}$ using the length scales discussed in Sec.~\ref{sec:length_scales}. For a trap configuration with $f_{r}=225$~Hz, $f_{L}=40$~Hz with 10 atoms per lattice site, the relevant separations are the mean interatomic distance $d\simeq 320$ nm within a lattice site, and the lattice spacing $D\simeq 407$ nm between adjacent sites in a Sr magic-wavelength optical lattice (Sec.~\ref{sec:length_scales}). 
\begin{figure}[h!]
  \resizebox{0.9\columnwidth}{!}{
  \includegraphics{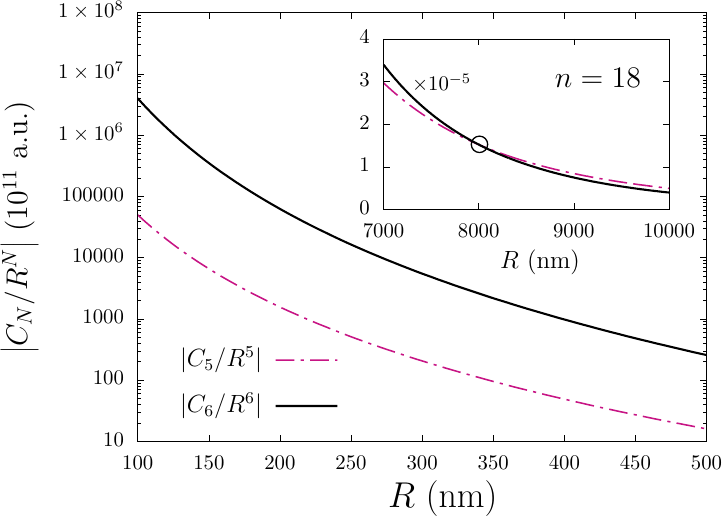}}
   \caption{Comparison of the quadrupole-quadrupole $E_{QQ}$ and van der Waals $E_{\rm vdW}$ interactions between two $5s18p\,{^1}P_1$ states of Sr as a function of the interatomic spacing. For the experimental parameters we consider in Sec.~\ref{sec:length_scales}, $|E_{\rm vdW}/E_{QQ}|\simeq 25$ for $n=18$. 
  }
  \label{fig:qq_vs_vdw}
\end{figure}

Fig.~\ref{fig:qq_vs_vdw} shows that, for two $5s18p\,{^1}P_1$ states, the van der Waals interaction dominates the quadrupole-quadrupole interaction over the length scales considered here. At the mean intra-site separation $d\simeq 320$ nm, $|E_{\rm vdW}/E_{QQ}|=|C_6/C_5|/R \simeq 25$ for $n=18$ with $\theta=\pi/2$. For atoms in adjacent lattice sites, separated by $D\simeq 407$ nm with $\theta=0$, the ratio remains about 12. Thus, the quadrupole-quadrupole interaction can be neglected for the singlet $P_1$ states in the low-$n$ range where the van der Waals interaction exhibits pronounced nonmonotonic behavior. The inset shows that, for the same-site geometry, the two interactions become comparable only at separations of order $8~\mu{\rm m}$.

\section{Conclusion}\label{sec:conc}
Motivated by the recent interest in using divalent Rydberg atoms in quantum information processing we have calculated the van der Waals coefficients for two interacting Sr Rydberg atoms.  We computed the $C_6$ coefficients for the van der Waals interactions between two $5sns({^1}S_0)$ and two $5snp({^1}P_{1,M_x=1})$ Sr atoms. We find that our results are in good agreement with previously reported values in Ref.~\cite{VaiJonPot12}, which tabulated these $C_6$ coefficients for $n\ge 30$. We also find that for $n<20$, the $C_6$ van der Waals coefficients for the $5snp({^1}P_1)$ states show strong non-monotonic deviations from the hydrogenic $n^{11}$ scaling due to the highly perturbed nature of the $5snd({^1}D_2)$ series as discussed in Ref.~\cite{VaiJonPot14}, which results in small energy denominators in the second-order expressions for the energy shifts. As a  result, the $C_6$ coefficients display highly non-monotonic behavior and change sign in the small $n$-region suggesting that the interaction can be made attractive or repulsive by choosing appropriate $n$.  Particularly at $n=18$, $\widetilde{C}_6$ is much larger than for any other state in the $n$-range we consider, opening the possibility of engineering strongly asymmetric long-range interactions by contrasting it with the van der Waals interaction between two Sr atoms in the $5sns({^1}S_0)$ states. The error bars in the low $n$-region obtained by propagating reported experimental  uncertainties  show that taking the experimental  uncertainties  into account does not change this picture for $n<20$.

\section{Acknowledgements}
This work was supported by the National Science Foundation (NSF) Grant No. PHY-1212482. A.D. was also supported by the Simons  Foundation  as a Simons fellow in theoretical physics, and by NSF Grant No. PHY-2513134. T.T. and A.D. would like to thank the Institute for Theoretical Atomic, Molecular, and Optical Physics (ITAMP) and the Harvard University Physics Department for their hospitality, where a part of this work was carried out. The authors would like to thank P. K{\'o}m{\'a}r and M. D. Lukin for valuable discussions.

\appendix
\begin{widetext}
  \section{Derivation of van der Waals expressions}\label{app:vdw_derivation}
  Here we derive an expression for the van der Waals interaction energy correction for a pair of atoms, $\delta E_{\rm vdW}$. The sets of quantum numbers of the two atoms (I and II) in the $LSJ$ coupling scheme will be denoted as $\gamma_1 [L_1 S_1 J_1]_{M_1}({\rm I})$ and $\gamma'_1 [L_1' S_1' J_1']_{M'_1}({\rm II})$, where the  quantum numbers describing the core  are denoted by $\gamma$, and $[L_1 S_1 J_1]_{M_1}$ and $[L_1' S_1' J_1']_{M'_1}$ refer to the Ry electrons with $n\ell$ and $n'\ell'$. 
  Since the quantization axis is directed along the lattice axis $\hat{x}$, the projections $M_1$ and $M_1'$ of the total angular momenta $J_1$ and $J_1'$ of the two atoms are $M_x$ and $M_x'$, respectively. The relative orientation of the quantization and internuclear axes is described by the angle $\theta$: $\theta=\pi/2$ for atom pairs within the same pancake-shaped cloud, and $\theta=0$ for atoms in different lattice sites along $\hat{x}$. 
  
  We are particularly interested in the van der Waals interactions between (1) two $5sns({^1}S_{0})$ atoms with  $M_x=M_x'=0$, and (2) $5snp({^1}P_{1,M_x})$ and $5snp({^1}P_{1,M_x'})$ atoms with $M_x=M_x'=1$. We keep our expressions in this appendix general enough without picking specific quantum numbers and $\theta$, the parameter controlling the angle between the internuclear and the quantization axes, for possible future implementations in different lattice geometries and choice of quantum states in quantum protocols where different $M$-states can become mixed (see for example~\cite{FiTsSu24}). 

  The second-order energy shift is given by
  \begin{align}\label{eq:denr}
    \delta E_{\rm vdW}
     & \big (\gamma_1[L_1 S_1 J_1]_{M_1} (\text{I}) \, \gamma'_1[L_1' S_1' J_1']_{M_1'}(\text{II}) \big ) =
    \frac{1}{R^6}                                                                                           \\
     & \times
    \sum_{\substack{{\gamma_2,\gamma_2'}                                                              \\ {J_2,\,J_2'}\\ {M_2,\,M_2'} }}
    \frac{
    \left | \left \bra \gamma_1[L_1 S_1 J_1]_{M_1}(\text{\small I})\, \gamma'_1[L_1' S_1' J_1']_{M_1'}(\text{\small II})
    \,\Big | V_{dd} \Big | \,
    \gamma_2[L_2 S_2 J_2]_{M_2}(\text{\small I})\, \gamma'_2[L_2' S_2' J_2']_{M_2'}(\text{\small II}) \right \ket \right |^2
    }{ \left (E_{\gamma_1[L_1 S_1 J_1]} + E_{\gamma_{1}'[L_1' S_1' J_1']} \right)
    - \left (E_{\gamma_2[L_2 S_2 J_2]} + E_{\gamma_2'[L_2' S_2' J_2']} \right) } \;, \notag
  \end{align}
  where  $V_{dd}$  is the rotated dipole-dipole interaction and the summation is over the intermediate states $\left| \gamma_2 [L_2 S_2 J_2]_{M_2}; \gamma'_2 [L_2' S_2' J_2']_{M_2'} \right\ket$. Because we are interested in the van der Waals interaction between atoms $\gamma_1 n \ell$ and $\gamma_1' n' \ell'$ in identical electronic configurations, we take  $\gamma_1=\gamma_1'$, $n = n'$ and $\ell = \ell'$. Intermediate states are $\gamma_2 n_2 \ell_2$ and $\gamma_2' n_2' \ell_2'$ in what follows where we also have $\gamma_2=\gamma_2'$. 

  To express the two-electron reduced matrix elements in terms of one-electron orbitals, we first transform the matrix element $\bra {^1}S_0 ||D|| {^1}P_1 \ket$ from the $LSJ$ to the $LS$ coupling,
  \begin{align} \label{eq:LSJ2LS}
    \bra \gamma_1[L_1 S_1 J_1] ||D|| \gamma_2 [L_2 S_2 J_2] \ket =
    (-1 & )^{L_1 +S_2 +J_2 +1} \sqrt{[J_1][J_2]} \\
        & \sixj{L_1}{J_1}{S_2}{J_2}{L_2}{1}
    \bra \gamma_1 L_1 ||D|| \gamma_2 L_2 \ket \delta_{S_1,S_2} \;, \notag
  \end{align}
  where $[L]=2L+1$. In the independent-particle approximation, the two-electron matrix element can be expressed in terms of a reduced matrix element involving only the single electron orbitals (see Ref.~\cite{JohPlaSap95} for details). Since we are only interested in the singlet states $S_{1}=S_{2}=0$, we obtain
  \begin{equation} \label{eq:c3_rmatel}
    \bra 5sn\ell(L_1) ||D|| 5sn'\ell'(L_2) \ket =
    \sqrt{[L_1][L_2]} (-1)^{\ell'+L_1 +1}
    \sixj{1}{L_2}{L_1}{0}{\ell}{\ell'}
    \bra n\ell ||d|| n'\ell' \ket \;.
  \end{equation}
  Here $d$ is the one-particle dipole operator and $D = \sum_{k=1}^2 d_{k}$ where the sum goes over the two atomic valence electrons. We also assume that the overlap between the $n\ell$ state of the Rydberg electron and the $5s$ state of the valence electron is negligible, which allows us to drop a second term involving $\bra 5s ||d|| n'\ell' \ket$ on the right-hand side of Eq.~\eqref{eq:c3_rmatel}. Since Eq.~\eqref{eq:c3_rmatel} now involves only the reduced matrix element for the Rydberg electron, we have removed the summation over the valence electrons and dropped the subscript in $d_{k}$.  Finally, we obtain
  \begin{equation}
     \begin{split}
      \delta E_{\rm vdW}
      \big ([L_1 &S_1 J_1]_{M_1} (\text{I}) \, [L_1' S_1' J_1']_{M_1'}(\text{II}) \big ) =
      \frac{1}{R^6}
      \sum_{\substack{{l_2,\,l_2'} \\{J_2,\,J_2'} }}  \\
      &\Big [
      \sum_{\substack{{M_1,\,M_1'}\\{M_2,\,M_2'}}}
      \sigma_{\theta}(J_1,\, J_1',\, M_1,\, M_1';\,J_2,\, J_2',\, M_2,\, M_2') \;
      \sigma_{\theta}(J_2,\, J_2',\, M_2,\, M_2';\,J_1,\, J_1',\, M_1,\, M_1')
      \Big ] \\
      &\times \bigg [ A(L_1,\, S_1,\, J_1,\,\ell_1;\, L_2,\, S_2,\, J_2,\, \ell_2)
        A(L_2',\, S_2',\, J_2',\, \ell_2';\, L_1',\, S_1',\, J_1',\,\ell_1') \bigg ]^2 \\
      &\times\sum_{n_2,n_2'} \frac{
        \Big | \big\bra n \ell \big| \big| d \big| \big| n_2 \ell_2 \big\ket \Big|^2 \;
        \Big | \big\bra n \ell \big| \big| d \big| \big| n_2' \ell_2' \big\ket \Big|^2
      }{\left (E_{L_1 S_1 J_1} + E_{L_1' S_1' J_1'} \right)
        - \left (E_{L_2 S_2 J_2} + E_{L_2' S_2' J_2'} \right) } \;. \label{eq:vdW_gen}
    \end{split}
  \end{equation}
  Here we have factored out the summation over the magnetic quantum numbers which depends on the rotation angle of the quantization axis. The function $\sigma_{\theta}$ describes the rotation of the quantization axis with respect to the inter-atomic axis, and the function $A$ involves the factors from breaking up the two-electron reduced matrix elements into reduced one-electron matrix elements in the LS coupling scheme. Explicitly, these functions are given by
  \begin{eqnarray}\nonumber
    \sigma_{\theta}(J_1,\, J_1',\, M_1,\, M_1';\,J_2,\, J_2',\, M_2,\, M_2') &=&
    \sum_{\mu} w^{(1)}_{\mu} (-1)^{M_1 + M_1'}
    \;d_{M_1-M_2, \mu}^{(1)}(\theta) \;d_{M_2'-M_1', -\mu}^{(1)}(\theta) \\ \nonumber
    &&\times\threej{J_1}{1}{J_2}{-M_1}{M_1-M_2}{M_2}
    \threej{J_1'}{1}{J_2'}{-M_1'}{-(M_2'-M_1')}{M_2'}\\ \nonumber
    A(L_1,\, S_1,\, J_1,\,l_1;\, L_2,\, S_2,\, J_2,\, l_2) &=&
    (-1)^{L_1 + S_2+ J_2 +l_2 + J_1 }
    \sqrt{[J_1][L_1][J_2][L_2]} \\
    && \times \sixj{1}{J_2}{J_1}{0}{l_1}{l_2}
    \sixj{L_1}{J_1}{S_2}{J_2}{L_2}{1} \;, \label{eq:sigma_a}
  \end{eqnarray}
  where $[J]=(2J+1)$ and ${d}^{(1)}_{M,\mu}(\theta)$ are Wigner functions. The angular dependence in Eq.~\eqref{eq:vdW_gen} can be expressed as a linear combination of Legendre polynomials $P_{0}(\cos\theta)$, $P_{2}(\cos\theta)$ and $P_{4}(\cos\theta)$. Finally, the $C_6$ coefficients can be extracted from this expression according to $\delta E_{\rm vdW}=-C_6/R^6$.

  For the vdW interaction between two atoms in the $5snp({^1}P_{1,1_{x}})$ states, the general expression~\eqref{eq:vdW_gen} can be broken into contributions from three channels: (1) $nP+nP\rightarrow n_2 S + n_2' S$, (2) $nP+nP\rightarrow n_2D+n_2'D$ and (3) $nP+nP\rightarrow n_2S+n_2'D$. Denoting the angular factors inside the closed brackets in Eq.~\eqref{eq:vdW_gen} with functions $f_{ss}(\theta)$, $f_{dd}(\theta)$ and $f_{sd}(\theta)$, we can express~\eqref{eq:vdW_gen} as 
  \begin{equation}\label{eq:vdw_angFunc}
    \delta E_{\rm vdW}  = -\frac{1}{R^6} \Big (
    f_{ss} (\theta) S_{ss} + f_{dd} (\theta) S_{dd} + f_{sd} (\theta) S_{sd}
    \Big ) \;,
  \end{equation}
  where the reduced sums $S_{ss}$, $S_{dd}$ and $S_{sd}$ are defined in Eq.~\eqref{eq:pp_red_sum}. \\

\end{widetext}

\section{Introducing singlet-triplet mixing into the $^{1}P_1+^{1}P_1$ interaction}\label{app:sing-trip}
We now describe how triplet character in the nominal singlet $D_2$ intermediate states is included in the $^{1}P_1+^{1}P_1$ calculation. The relevant matrix elements in Eq.~\eqref{eq:denr} are 

\begin{align}
  & \left\langle {^1}P_{1}\text { (I) }{ } {^1}P_{1} \text { (II) }\left|V_{DD}\right| {^1}S_0 \text { (I) } {^1}S_0 \text { (II) }\right\rangle \\
  & \left\langle {^1}P_{1}\text { (I) }{ } {^1} P_{1} \text { (II) }\left|V_{DD}\right| {^1}S_0 \text { (I) } {^1}D_2 \text { (II) }\right\rangle \\
  & \left\langle {^1}P_{1}\text { (I) } {^1}P_{1} \text { (II) }\left|V_{DD}\right| {^1}D_2\text { (I) } {^1}S_0\text { (II) }\right\rangle      \\
  & \left\langle {^1}P_{1}\text { (I) }{ }^{1} P_{1} \text { (II) }\left|V_{DD}\right|{^1} D_2 \text { (I) }{ }{^1} D_2 \text { (II) }\right\rangle \\
  & \left\langle {^1}P_{1}\text { (I) } {^1}P_{ 1} \text { (II) }\left|V_{DD}\right|{ }{^1}D_2\text { (I) } {^3}D_2\text { (II) }\right\rangle  \\
  & \left\langle {^1}P_{1}\text { (I) } {^1}P_{ 1} \text { (II) }\left|V_{DD}\right|{ }{^3}D_2\text { (I) } {^1}D_2\text { (II) }\right\rangle  \\
  & \left\langle {^1}P_{1}\text { (I) } {^1}P_{1}\text { (II) }\left|V_{DD}\right|{^3}D_2\text { (I) } {^3}D_2 \text { (II) }\right\rangle  \\
  & \left\langle {^1}P_{1}\text { (I) }{ }^{1}P_{1}\text { (II) }\left|V_{DD}\right| {^1}S_0\text { (I) }{ }{^3}D_2 \text { (II) }\right\rangle  \\
  & \left\langle {^1}P_{1}\text { (I) }{ }^{1}P_{1}\text { (II) }\left|V_{DD}\right| { }{^3}D_2\text { (I) } {^1}S_0 \text { (II) }\right\rangle              
\end{align}

We include triplet admixture in the nominal singlet $D_2$ intermediate states by making the replacement 
\begin{equation*}
  \left|^{1}D_2\right\rangle \rightarrow \alpha_1\left|^{1}D^*_2\right\rangle+\alpha_3\left|{ }^3 D^*_2\right\rangle \;.
\end{equation*}

%
The terms ${^1}D_2^*$ and ${ }^3 D_2^*$ here label the singlet and triplet characters in the ${^1}D_2$ states. The matrix elements B2 and B3, B5 and B6, and B8 and B9 are equal, and there is no triplet mixing into the $^{1}S_0$ states. As an example, we will investigate the effect of the singlet-triplet mixing on the matrix element B4:
\begin{equation}\label{eq:dd-matel}
  \begin{split}
     & \left\langle {^1}P_{1} \text { (I) } {^1}P_{1} \text { (II) }\left|V_{DD}\right| {^1}D_2 \text { (I) } {^1}D_2 \text { (II) }\right\rangle \\
     & \begin{aligned}
      =-\frac{1}{R^3}  \sum_\mu &w_\mu^{(1)} \left\langle {^1}P_{1}\text { (I) }\left|D_\mu^{(1)}(\text {I}, \hat{S} \hat{\mathbf{r}})\right| {^1}D_2\text { (II) }\right\rangle  \\
      & \times\left\langle {^1}P_{1}(\text{I})\left|D_{-\mu}^{(1)}(\text {II}, \hat{S} \hat{\mathbf{r}})\right| {^1}D_2\text { (II) }\right\rangle
    \end{aligned}
  \end{split}
\end{equation}

Here $D_\mu^{(1)}(\hat{S} \hat{\mathbf{r}})$ are the dipole operators rotated according to Eq.~\eqref{eq:d1_rotated} so that the internuclear axis is the quantization axis. With the replacement $\left|{^1}D_2\right\rangle \rightarrow \alpha_1\left|{^1}D_2^*\right\rangle+\alpha_3\left|{ }^3 D_2^*\right\rangle$, the numerator in the van der Waals interaction energy $\delta E_{vdW}$ for the ${^1}D_2$--${^1}D_2$ channel becomes

\begin{widetext}

  \begin{equation}
    \begin{aligned}
      \langle {^1}P_{1} & \text { (I) } {^1}P_{1} \text { (II) }\left|V_{DD}\right|{ }^{1} D_2\text { (I) } {^1}D_2 \text { (II) }\rangle                     \\
        & \left\langle{ }^{1} D_2\text { (I) } {^1}D_2 \text { (II) }|V_{DD}|{ }^{1} P_{1}\text { (I) } {^1}P_{1}\text { (II) }\rangle\right. \\
        & \begin{aligned}
        =\biggl(-\frac{1}{R^3} \sum_\mu w_\mu^{(1)} \left[ {\alpha_1}\langle {^1}P_{1}\text { (I) }|D_\mu^{(1)}(\text{I}, \hat{S} \hat{\mathbf{r}})|^1 D_2^*\text { (I) }\rangle + {\alpha_3}{\langle { }^{1}P_{1}\text { (I) }|D_\mu^{(1)}(\text{I}, \hat{S} \hat{\mathbf{r}})|{ }^3 D_2^*\text { (I) }\rangle}\right] \\
        \times \left[ {\alpha_1}\langle {^1}P_{1} \text { (II) }|D_{-\mu}^{(1)}(\text{II}, \hat{S} \hat{\mathbf{r}})|^1 D_2^*\text { (II) }\rangle
        + {\alpha_3}{\langle{ }^{1} P_{1}\text { (II) }|D_{-\mu}^{(1)}(\text{II}, \hat{S} \hat{\mathbf{r}})|{ }^3 D_2^*\text { (II) }\rangle} \right] \biggr)
      \end{aligned}                                                                                                          \\
        & \begin{aligned}
        \times\biggl(-\frac{1}{R^3} \sum_\mu w_\mu^{(1)} \left[ {\alpha_1^*}\langle ^1 D_2^*\text { (I) } |D_\mu^{(1)}(\text{I}, \hat{S} \hat{\mathbf{r}})|{^1}P_{1}\text { (I) }\rangle + {\alpha_3^*} {\langle { }^3 D_2^*\text { (I) }|D_\mu^{(1)}(\text{I}, \hat{S} \hat{\mathbf{r}})|{ }^{1}P_{1}\text { (I) }\rangle}\right] \\
        \times \left[ {\alpha_1^*}\langle ^1 D_2^*\text { (II) }|D_{-\mu}^{(1)}(\text{II}, \hat{S} \hat{\mathbf{r}})|{^1}P_{1} \text { (II) }\rangle
        + {\alpha_3^*} {\langle { }^3 D_2^*\text { (II) }|D_{-\mu}^{(1)}(\text{II}, \hat{S} \hat{\mathbf{r}})|{ }^{1} P_{1}\text { (II) }\rangle} \right] \biggr)
      \end{aligned}                                                                                                          \\
       &
    \end{aligned}
  \end{equation}

\end{widetext}


The matrix elements multiplying $\alpha_3$ and $\alpha_3^*$ connect the ${^1}P_1-{ }^3 D_2^*$ states, and they are exactly zero in the nonrelativistic formalism we adopt in this paper. The van der Waals interaction energy can be split up in terms of the matrix elements involving the singlet-singlet, singlet-triplet, and triplet-triplet channels. The procedure for the ${^1}D_2$--${^1}D_2$ matrix element above can be repeated for each of the matrix elements involving ${^1}D_2$ and ${^3}D_2$ states as intermediate states. We replace $\left|{^1}D_2\right\rangle \rightarrow \alpha_1\left|{^1}D_2^*\right\rangle+\alpha_3\left|{ }^3 D_2^*\right\rangle$ and $\left.\left|{ }^3 D_2\right\rangle \rightarrow \beta_1\left|{ }^1 D_2^*\right\rangle+\left.\beta_3\right|^3 D_2^*\right\rangle$ in these intermediate states. The terms ${^1}D_2^*$ and ${ }^3 D_2^*$ again label the actual singlet/triplet characters in the broken LSJ-coupling of ${^1}D_2$ and ${ }^3 D_2$ states according to Eq.~\eqref{eq:dd-matel}. The energy denominators remain unaffected. Ignoring the relativistically suppressed singlet-triplet matrix elements as before, we obtain

\begin{widetext}
  \begin{equation}\label{eq:sing-trip-split}\displaystyle
    \begin{aligned}
      \delta E_{v d W}=\frac{1}{R^6} \sum_{n_2,n_2^{\prime}}\;\;  &\frac{\left|\left\langle\ {^1}P_1\text { (I)}{ }^{1} P_1\text { (II) }\left|V_{DD}\right| n_2 {^1}S_0(I){\;}n_2'{^1} S_0 \text { (II) }\right\rangle\right|^ 2}{2 E_{{^1}P_1}-\left(E_{{^1}S_0}+E_{{^1}S_0}\right)}  \\
      & +2\,|\alpha_1|^2\, \frac{\left|\left\langle {^1}P_1 {^1}P_1\left| V_{DD} \right| n_2 {^1}S_0 \; n_2' {^1}D^*_2\right\rangle\right|^2}{2 E_{{^1}P_1}-\left(E_{{^1}S_0}+E_{{^1}D_2}\right)}  
       +|\alpha_1|^4\, \frac{\left|\left\langle {^1}P_{ 1 } { } {^1}P_{ 1 } \left|V_{DD}\right|{ }n_2{^1} D^*_2{ }\; n_2'{^1}D^*_2\right\rangle\right|^ 2}{2 E_{{^1}P_1}-\left(E_{{^1}D_2}+E_{{^1}D_2}\right)}  \\
      & + 2\,|\alpha_1|^2 |\beta_1|^2 \,\frac{|\left\langle {^1}P_{1} {^1}P_{ 1 } \left|V_{DD}\right|n_2{^1}D^*_2 \; n_2'{^1}D^*_2\right\rangle|^2}{2 E_{{^1}P_1}-\left(E_{{^1}D_2} + E_{{ }{^3}D_2}\right)} \\
      & + |\beta_1|^4 \,\frac{\left\langle {^1}P_1 {^1}P_1 \left|V_{DD}\right| n_2{^1}D^*_2 \; n_2'{^1}D^*_2 \right\rangle|^2}{2 E_{{^1}P_1}-\left(E_{{ }{^3} D_2}+E_{{^3}D_2}\right)}  
       +2\, |\beta_1|^2 \, \frac{\left\langle {^1}P_1{ }{^1}P_1\left|V_{DD}\right| n_2{^1}S_0 \; n_2'{^1}D^*_2\rangle\right|^2}{2 E_{{^1}P_1}-\left(E_{{^1}S_0}+E_{{^3}D_2}\right)} \,.
    \end{aligned}
  \end{equation}
The sum is over the principal quantum numbers $n_2$, $n_2'$ of the intermediate states. We dropped the labels I and II after the first term in the sum for brevity. The coefficients $\alpha_1$ and $\beta_1$ are the channel fractions quoted in Fig. 7 (a) and (c) of~\cite{VaiJonPot14}.  

Generally, $\Psi_{J M}$  eigenstates of interest may be expanded as,  
\begin{equation*}
  \Psi_{J M}=\sum_{i<j} c_{i j} \Phi_{i j} \,,
\end{equation*}
where the quantities $c_{i j}$ are expansion coefficients and determine the admixture of each channel wave function $\Phi_{i j}$ in the eigenstate. The two-electron reduced matrix elements in Eq.~\eqref{eq:LSJ2LS} are expressed in terms of one-electron reduced matrix elements and the coefficients $c_{ij}$ through the standard decomposition of Ref.~\cite{JOHNSON1995255} where the channel fractions reported in Ref.~\cite{VaiJonPot14} correspond to $|c_{ij}|^2$:

  \begin{equation}
    \begin{aligned}
       & \left\langle j_r j_s\left(J_F\right)\left\|T_J\right\| j_m j_n\left(J_I\right)\right\rangle=\sqrt{\left(2 J_I+1\right)\left(2 J_F+1\right)}(-1)^J \sum_{\substack{m \leq n                                                                                      \\ r \leq s}} \eta_{r s} \eta_{m n} c_{r s} c_{m n} \\
       & \times\left[(-1)^{j_r+j_s+J_I}\left\{\begin{array}{ccc}
          J   & J_I & J_F \\
          j_s & j_r & j_m
        \end{array}\right\}\left\langle r\left\|t_J\right\| m\right\rangle \delta_{n s}+(-1)^{j_r+j_n}\left\{\begin{array}{lll}
          J   & J_I & J_F \\
          j_s & j_r & j_n
        \end{array}\right\}\left\langle r\left\|t_J\right\| n\right\rangle \delta_{m s}\right.  \\
       & \left.+(-1)^{J_F+J_i+1}\left\{\begin{array}{ccc}
          J   & J_I & J_F \\
          j_r & j_s & j_m
        \end{array}\right\}\left\langle s\left\|t_J\right\| m\right\rangle \delta_{n r}+(-1)^{j_r+j_n+J_F}\left\{\begin{array}{lll}
          J   & J_I & J_F \\
          j_r & j_s & j_n
        \end{array}\right\}\left\langle s\left\|t_J\right\| n\right\rangle \delta_{m r}\right] \;, 
    \end{aligned}
  \end{equation}
  where only the last term survives in our case. The symbols $\eta_{ij}$ are normalization factors and are given by 
  $$
  \eta_{i j}= \begin{cases}1, & i \neq j \\ 1 / \sqrt{2}, & i=j .\end{cases}
  $$
  From Eq.~\eqref{eq:sing-trip-split}, we can define reduced sums $S_{ss}$, $S_{dd}$, and $S_{sd}$ that involve only the radial matrix elements and the energy denominators by factoring out the angular dependencies in Eq.~\eqref{eq:vdW_gen}:

  \begin{equation}\label{eq:pp_red_sum}
    \begin{split}
      S_{ss} =&
      -\sum_{\substack{{n_2,n_2'} }}
      \frac{
        \Big | \big\bra np \big| \big| d^{(1)} \big| \big| n_2 s \big\ket \Big|^2 \;
        \Big | \big\bra np \big| \big| d^{(1)} \big| \big| n_2' s \big\ket \Big|^2
      }{ 2E_{np\, {^1}\!P_1}
        - \left (E_{n_2 s\, {^1}\!S_0} + E_{n_2' s\, {^1}\!S_0} \right) } \, , \\ 
      S_{dd} =&
      -\sum_{\substack{{n_2,n_2'} }}
      |\alpha_1|^4 \frac{
        \Big | \big\bra np \big| \big| d^{(1)} \big| \big| n_2 d \big\ket \Big|^2 \;
        \Big | \big\bra np \big| \big| d^{(1)} \big| \big| n_2' d \big\ket \Big|^2
      }{ 2E_{np\, {^1}\!P_1}
        - \left (E_{n_2 d\, {^1}\!D_2} + E_{n_2' d\, {^1}\!D_2} \right) }
      + 2|\alpha_1|^2 |\beta_1|^2 \frac{
        \Big | \big\bra np \big| \big| d^{(1)} \big| \big| n_2 d \big\ket \Big|^2 \;
        \Big | \big\bra np \big| \big| d^{(1)} \big| \big| n_2' d \big\ket \Big|^2
      }{ 2E_{np\, {^1}\!P_1}
        - \left (E_{n_2 d\, {^1}\!D_2} + E_{n_2' d\, {^3}\!D_2} \right) } \\
      &+ |\beta_1|^4 \frac{
        \Big | \big\bra np \big| \big| d^{(1)} \big| \big| n_2 d \big\ket \Big|^2 \;
        \Big | \big\bra np \big| \big| d^{(1)} \big| \big| n_2' d \big\ket \Big|^2
      }{ 2E_{np\, {^1}\!P_1}
        - \left (E_{n_2 d\, {^3}\!D_2} + E_{n_2' d\, {^3}\!D_2} \right) } \, ,\\
      S_{sd} =&
      -\sum_{\substack{{n_2,n_2'} }}
      2|\alpha_1|^2 \frac{
        \Big | \big\bra np \big| \big| d^{(1)} \big| \big| n_2 s \big\ket \Big|^2 \;
        \Big | \big\bra np \big| \big| d^{(1)} \big| \big| n_2' d \big\ket \Big|^2
      }{ 2E_{np\, {^1}\!P_1}
        - \left (E_{n_2 s\, {^1}\!S_0} + E_{n_2' d\, {^1}\!D_2} \right) } 
      + 2|\beta_1|^2 \frac{
        \Big | \big\bra np \big| \big| d^{(1)} \big| \big| n_2 s \big\ket \Big|^2 \;
        \Big | \big\bra np \big| \big| d^{(1)} \big| \big| n_2' d \big\ket \Big|^2
      }{ 2E_{np\, {^1}\!P_1}
        - \left (E_{n_2 s\, {^1}\!S_0} + E_{n_2' d\, {^3}\!D_2} \right) }\;. \\
    \end{split}
  \end{equation}

  With these definitions of the reduced sums, we can express the $C_6$ coefficients as in Eq.~\eqref{eq:vdw_angFunc} where the angle $\theta$ between the internuclear and the optical lattice axes is explicitly spelled out. \\

\section{Derivations of the Quadrupole-Quadrupole interactions}\label{sec:qq_derivation}

To calculate the Quadrupole-Quadrupole interaction between two $\left|{^1}P_{1, M}\right\rangle$ states, we start with the two-atom state 
\begin{equation}
  |\psi\rangle=\frac{1}{\sqrt{2}} \Big[ \left|{^1}P_{1, M_1}\right\rangle_{\text {I}}\left|{^1}P_{1, M_2}\right\rangle_{\text {II }} \pm \left|{^1}P_{1, M_2}\right\rangle_{\text {I}}\left|{^1}P_{1, M_1}\right\rangle_{\mathbb{II}} \Big]
\end{equation}

The quadrupole tensor is given by $Q_\mu^{(2)}=-|e| \sum_j r_j^2 C_\mu^{(2)}\left(\hat{\mathbf{r}}_j\right)$, where $C_\mu^{(k)}=\sqrt{4\pi/(2 k+1)} Y_\mu^{(k)}$ are the normalized spherical harmonics. The Quadrupole-Quadrupole interaction in terms of the Quadrupole tensor reads  
\begin{equation}
  V_{Q Q}=\frac{1}{R^5} \sum_\mu \frac{4 !}{(2-\mu) !(2+\mu) !} Q_\mu^{(2)}({\text {I},\hat{S} \hat{\mathbf{r}}}) Q_{-\mu}^{(2)}({\text {II}},\hat{S} \hat{\mathbf{r}})\, , \\
\end{equation}
where we have rotated the quantization axis as before, $\hat{\mathbf{r}} \rightarrow \hat{S} \hat{\mathbf{r}}$. For the rest of this discussion, we will denote the two-atom product states as $\left|{^1}P_{1, M_1}\right\rangle_{\text {I}}\left|{^1}P_{1, M_2}\right\rangle_{\mathbb{II}} := \left|{^1}P_{1, M_1}({\text {I}}) {^1}P_{1, M_2}({\text {II}})\right\rangle$ for brevity. The matrix elements become  
\begin{equation}
  \begin{aligned}
    E_{QQ} = \frac{1}{2} &\Big[\left\langle {^1}P_{1, M_1}({\text {I}}) {^1}P_{1, M_2}({\text {II}}) \right| \pm \left\langle {^1}P_{1, M_1}({\text {I}}) {^1}P_{1, M_2}({\text {II}}) \right| \Big]   \\
    &\times V_{QQ} \Big[ \left|{^1}P_{1, M_1^{\prime}}({\text {I}}) {^1}P_{1, M_2^{\prime}}({\text {II}}) \right\rangle  \pm \left|{^1}P_{1, M_1^{\prime}}({\text {I}}) {^1}P_{1, M_2^{\prime}}({\text {II}}) \right\rangle \Big]\, . \\
  \end{aligned}
\end{equation}

$E_{QQ}$ can be expressed in terms of the individual two-atom matrix elements,
\begin{equation}
  \begin{aligned}
    E_{QQ} = & \frac{1}{2} \Big[\left\langle {^1}P_{1, M_1}\text { (I) } {^1}P_{1, M_2} \text { (II) }\left|V_{Q Q}\right| {^1}P_{1, M_1^{\prime}}\text { (I) } {^1}P_{1, M_2^{\prime}}\text { (II) }\right\rangle  \\ 
    & \pm \left\langle {^1}P_{1, M_1}\text { (I) } {^1}P_{1, M_2}\text { (II) }\left|V_{Q Q}\right| {^1}P_{1, M_2^{\prime}}\text { (I) } {^1}P_{1, M_1^{\prime}} \text { (II) }\right\rangle \\ 
    & \pm\left\langle {^1}P_{1, M_2} \text { (I) } {^1}P_{1, M_1} \text { (II) }\left|V_{Q Q}\right| {^1}P_{1, M_1^{\prime}} \text { (I) }{ } {^1}P_{1, M_2^{\prime}} \text { (II) }\right\rangle \\ 
    &  +\left\langle {^1}P_{1, M_2} \text { (I) } {^1}P_{1, M_1}\text { (II) } \left| V_{Q Q}\right|{ } {^1}P_{1, M_2^{\prime}}\text { (I) } {^1}P_{1, M_1^{\prime}}\text { (II) } \right\rangle \Big] \, . \\
  \end{aligned}
\end{equation}

All four of these matrix elements are equal, and therefore only the {\it gerade} states will have nonzero quadrupole interactions. Thereby, we only consider the {\it gerade} states: 
\begin{equation}
  \begin{aligned}
    \langle {^1}P_{1, M_1}&\text {(I)} {^1}P_{1, M_2} \text {(II)}\left|V_{Q Q}\right| {^1}P_{1, M_1^{\prime}}\text {(I)} {^1}P_{1, M_2^{\prime}}\text {(II)}\rangle \\ 
       &= \frac{1}{R^5} \sum_\mu \frac{4 !}{(2-\mu) !(2+\mu) !}\left\langle {^1}P_{1, M_1}\text {(I)} \left|Q_\mu^{(2)}(\text {I}, \hat{S} \hat{\mathbf{r}})\right| {^1}P_{1, M_1^\prime}\text {(I)}\right\rangle \\ 
      &\times \left\langle {^1}P_{1, M_2}\text {(II)} \left|Q_{-\mu}^{(2)}(\text {II}, \hat{S} \hat{\mathbf{r}})\right| {^1}P_{1, M_2^\prime}\text {(II)}\right\rangle \,.
  \end{aligned}
\end{equation}
Here rotated quadrupole tensors are given by 
\begin{equation}\label{eq:q2_rotated}
  Q_\mu^{(2)}(\hat{S} \hat{\mathbf{r}})=\sum_{m=-2}^2 Q_m^{(2)}(\hat{\mathbf{r}}) \mathbb{D}_{m, \mu}^{(2)}(0, \theta, 0)\, ,
\end{equation}
where $\mathbb{D}_{m, \mu}^{(2)}$ are the Wigner functions. In terms of the unrotated Quadrupole tensors, the matrix element becomes, 
\begin{equation}
  \begin{aligned}
  \langle {^1}P_{1, M_1}&\text {(I)} {^1}P_{1, M_2} \text {(II)}\left|V_{Q Q}\right| {^1}P_{1, M_1^{\prime}}\text {(I)} {^1}P_{1, M_2^{\prime}}\text {(II)}\rangle \\ 
  & = \frac{1}{R^5} \sum_\mu \frac{4 !}{(2-\mu) !(2+\mu) !} \sum_{m, m^{\prime}} \mathbb{D}_{m, \mu}^{(2)}(0, \theta, 0)\left\langle {^1}P_{1, M_1}\text {(I)} \left|Q_m^{(2)}(\text {I}, \hat{\mathbf{r}})\right| {^1}P_{1, M_1^\prime}\text {(I)} \right\rangle \\
  & \times \mathbb{D}_{m^{\prime},-\mu}^{(2)}(0, \theta, 0)\left\langle {^1}P_{1, M_2}\text {(II)} \left|Q_{m^\prime}^{(2)}(\text {II}, \hat{\mathbf{r}})\right| {^1}P_{1, M_2^\prime}\text {(II)} \right\rangle \\
  &=\frac{(-1)^{M_1+M_2}}{R^5} \sum_\mu \frac{4 !}{(2-\mu) !(2+\mu) !} \mathbb{D}_{M_1-M_1^{\prime}, \mu}^{(2)}(0, \theta, 0) \mathbb{D}_{M_2-M_2^{\prime},-\mu}^{(2)}(0, \theta, 0) \\ & \times\left(\begin{array}{ccc}1 & 2 & 1 \\ -M_1 & \left(M_1-M_1^{\prime}\right) & M_1^{\prime}\end{array}\right)\left(\begin{array}{ccc}1 & 2 & 1 \\ -M_2 & \left(M_2-M_2^{\prime}\right) & M_2^{\prime}\end{array}\right) \left|\left\langle {^1}P_1\left\|Q^{(2)}\right\| {^1}P_1\right\rangle\right|^2 
  \end{aligned}
\end{equation}
where in the last step, we employed the Wigner-Eckart theorem. The sum over $\mu$ and the Clebsch-Gordan coefficients can be collected into one function 
\begin{equation}
  \begin{aligned}
    S_Q\left(M_1, M_1^{\prime}, M_2, M_2^{\prime} ; \theta\right)=&(-1)^{M_1+M_2}\left(\begin{array}{ccc}1 & 2 & 1 \\ -M_1 & \left(M_1-M_1^{\prime}\right) & M_1^{\prime}\end{array}\right) \left(\begin{array}{ccc}1 & 2 & 1 \\ -M_2 & \left(M_2-M_2^{\prime}\right) & M_2^{\prime}\end{array}\right) \\ 
    & \times \sum_{\mu=-2}^2 \frac{4 !}{(2-\mu) !(2+\mu) !} \mathbb{D}_{M_1-M_1^{\prime}, \mu}^{(2)}(0, \theta, 0) \mathbb{D}_{M_2-M_2^{\prime},-\mu}^{(2)}(0, \theta, 0)\, , 
  \end{aligned}
\end{equation}
which simplifies the matrix element: 
\begin{equation}
  \begin{aligned}
    \langle {^1}P_{1, M_1}&\text {(I)} {^1}P_{1, M_2} \text {(II)}\left|V_{Q Q}\right| {^1}P_{1, M_1^{\prime}}\text {(I)} {^1}P_{1, M_2^{\prime}}\text {(II)}\rangle \\
    = &\frac{1}{R^5} S_Q\left(M_1, M_1^{\prime}, M_2, M_2^{\prime} ; \theta\right)\left|\left\langle {^1}P_1\left\|Q^{(2)}\right\| {^1}P_1\right\rangle\right|^2 \, .
  \end{aligned}
\end{equation}

When $M_1=M_1^{\prime}=1$ and $M_2=M_2^{\prime}=1$ and the quantization axis is along the axis of the optical lattice, $\theta=\pi / 2$, $S_Q(1,1,1,1 ; \pi/2)=3/40$, and the matrix element becomes 
\begin{equation}
  \begin{aligned}
    \langle {^1}P_{1, M_1} \text {(I)} {^1}P_{1, M_2} \text {(II)}\left|V_{Q Q}\right| {^1}P_{1, M_1^{\prime}}\text {(I)} {^1}P_{1, M_2^{\prime}}\text {(II)}\rangle 
    = \frac{3}{40 R^5} \left|\left\langle {^1}P_1\left\|Q^{(2)}\right\| {^1}P_1\right\rangle\right|^2 \, .
  \end{aligned}
\end{equation}

The two-electron reduced matrix element $\left\langle {^1}P_1\left\|Q^{(2)}\right\| {^1}P_1\right\rangle$ can be expressed in terms of the single-electron reduced matrix element $\left\langle n p\left\|q^{(2)}\right\| n p\right\rangle$~\cite{JOHNSON1995255}, and we obtain 
\begin{equation}\label{eq:eq_one_el_mel}
  \begin{aligned}
    E_{QQ} &= 2\langle {^1}P_{1, M_1} \text {(I)} {^1}P_{1, M_2} \text {(II)}\left|V_{Q Q}\right| {^1}P_{1, M_1^{\prime}}\text {(I)} {^1}P_{1, M_2^{\prime}}\text {(II)}\rangle \\
    &= \frac{3}{20 R^5} \left|\left\langle n p\left\|q^{(2)}\right\| n p\right\rangle\right|^2 \, .
  \end{aligned}
\end{equation}

The one-electron reduced matrix element in Eq.~\ref{eq:eq_one_el_mel} scales as $n^4$ after $n\sim 8$, and it converges to $\left|\left\langle n p\left\|q^{(2)}\right\| n p\right\rangle\right|/n^4\sim 2.7$ for $n>8$. This gives us a final expression: $E_{QQ} = \widetilde{C}_5 n^8 /R^5$ where $\widetilde{C}_5\approx 1.1$ and $C_5=\widetilde{C}_5 \,n^8$.

\end{widetext}

\section{Error estimates for the dipole matrix elements}\label{app:matel_err}
Here we show how big the errors in our calculated matrix elements need to be to increase the sizes of the error bars in Fig.~\ref{fig:C6_coeff_pp}, effectively recovering the $n^{11}$ non-monotonic scaling (blue diamonds in Fig.~\ref{fig:C6_coeff_pp}). Specifically, we conclude that to change the size of the error bar for $n=18$ so that it becomes large enough to encapsulate its monotonically scaling counterpart (blue diamond at $n=18$), the error in the corresponding radial matrix element needs to be much larger than what is reported in~\cite{VaiJonPot12}. Therefore, we conclude that using the model potential does not alter our qualitative arguments.

In Sec.~\ref{subsec:c6_err}, we estimated the uncertainty in the $C_6$ coefficients under the assumption that the uncertainty in the dipole matrix elements was zero, {\it i.e.}, $\delta(V_{j,k})=0$ (Eq.~\eqref{eq:err_quadrature}). Instead, we now start with the fractional errors~\cite{Taylor1997}. Based on Fig.~\ref{fig:EnrDominator}, we will assume that the sum in Eq.~\eqref{eq:vdW_r6} is dominated by a single term, $nP+nP\rightarrow nD+(n-1)D$ where $n=18$. In the single-electron, single-term approximation, there are two distinct matrix elements in $V_{j,k}$, which we label $d_{n,n_2}$ and $d_{n,n_2'}$, and we remove the sum. These matrix elements are then squared in the numerator of~\eqref{eq:vdW_r6}. Therefore, the fractional error is
\begin{equation}\label{eq:frac_err_c6}
  \left( \frac{\delta C_{6}}{C_6}\right )^2 = 2 \left[ \left(\frac{\delta d_{n,n_2}}{d_{n,n_2}}  \right)^2 + \left( \frac{\delta d_{n,n_2'}}{d_{n,n_2'}}  \right)^2 \right] + \left( \frac{\delta \Delta E}{\Delta E} \right)^2 \;.
\end{equation}
%
Here we added the fractional uncertainties in quadrature, treating the uncertainties in the energies and in the dipole matrix elements as independent. The experimental energy uncertainties are themselves mostly correlated, as they come from a single experimental group~\cite{SanNav10}; however, summing them linearly rather than in quadrature changes the calculated error bars in Fig.~\ref{fig:C6_coeff_pp} negligibly. In estimating the uncertainties in Sec.~\ref{subsec:c6_err}, we assumed that $\delta d_{n,n_2}=\delta d_{n,n_2'}=0$. This assumption leads to
\begin{equation}
  \frac{\delta C^{(0)}_{6}}{C^{(0)}_6} \coloneqq \frac{\delta C_{6}}{C_6} = \frac{\delta \Delta E}{\Delta E} \,.
\end{equation}
From Table~\ref{Table:errBars}, $\delta C^{(0)}_{6}/C^{(0)}_{6} = 0.44$. When $n=18$ and $n_2'=17$, the matrix elements $d_{n,n}$ and $d_{n,n-1}$ scale as $n^2$ with similar coefficients~\cite{HoNo98} which cancel from the fractional errors, and $\delta d_{n,n}/d_{n,n} \approx \delta d_{n,n-1}/d_{n,n-1}$. This allows us to write from~\eqref{eq:frac_err_c6}
\begin{equation}\label{eq:frac_err_d_nu}
  \left( \frac{\delta d_{n,n}}{d_{n,n}} \right)^2 = \frac{1}{4} \left[ \left( \frac{\delta C^{(1)}_{6}}{C^{(1)}_6} \right)^2 - \left( \frac{\delta C^{(0)}_{6}}{C^{(0)}_6} \right)^2 \right] \;,
\end{equation}
where $C^{(1)}_6$ is the actual value of the $C_6$ coefficient that takes the uncertainties in the dipole matrix elements into account. For the error bars in Fig.~\ref{fig:C6_coeff_pp} to become large enough to encapsulate the $C_6$ coefficients obtained using the model potential alone (blue diamonds), $\delta C^{(1)}_{6}/C^{(1)}_6 \approx 0.98$, and $\delta d_{18,18}/d_{18,18} \approx \delta d_{18,17}/d_{18,17} \approx 0.44$.
Furthermore, since $d_{n,n} \propto (n^*)^2 = (n-\nu_{n})^2$, and the quantum defects $\nu_{n}$ change little as a function of $n$~\cite{HoNo98}, we have
\begin{equation}\label{eq:frac_err_d}
   \frac{\delta d_{n,n}}{d_{n,n}} =  \frac{2\delta\nu_{n}}{(n-\nu_{n})} \approx 0.44 \;.
\end{equation}
For $n=18$, therefore $\delta(\nu_{n})\approx 3.35$. This value can then be regarded as the uncertainty in the quantum defect the  $n=18$ states need to have for the corresponding dipole matrix element to become large enough to undo the non-monotonic resonance-like behavior we discuss in this work. 
We note that Eq.~\eqref{eq:frac_err_d} also accounts for the propagation of the state-energy errors into the radial matrix elements: an error in the energy of a Rydberg state is equivalent to an error $\delta(\nu_n)$ in its quantum defect, since $\nu_n$ is fixed by the level energy, and this error propagates into the matrix elements through the $(n^*)^2$ scaling. In the sensitivity analysis below, we will examine this propagation more directly by varying the model potential parameters, which vary the calculated state energies and the radial matrix elements together. 

We first note that the value $\delta(\nu_n)\approx 3.35$ is too large even for a model potential. For comparison, the quantum defect $\nu_n$ varies by 0.55 across the $^1S_0$ and $^1P_1$ Ry series, and by 0.34 between the $^1P_1$ and $^1D_2$ series of Sr in the $n$-range of interest. Similarly, across the triplet Ry series, $\nu_n$ varies by 0.47 between the $^3S_1$ and the $^3P_{0,1,2}$ series, and by 0.3 between the $^3P_{0,1,2}$ and $^3D_{1,2,3}$. Even across different species, for example between Sr and Ca, $\nu_n$ varies by 0.8 across the 5s$n$p$(^1P_1)$ and 4s$n$p$(^1P_1)$. 

This suggests that $\delta(\nu_{18})\approx 3.35$ is too large an error to expect in the quantum defect. Therefore, we expect that using a model potential for evaluating the radial wave functions  is justified in our calculations.
For example, the quantum defect for the $5s18p({^1}P_{1})$ state obtained in~\cite{VaiJonPot12} by fitting to the Rydberg-Ritz formula~\eqref{eq:ry-ritz} using experimental data is 2.71. On the other hand, our model potential yields a quantum defect of 2.33. The difference is chiefly due to two factors: (1) the model potential in~\cite{Millen11} fits  experimental energies for $n>20$ where the quantum defects vary smoothly, and (2) our fit targets the energy denominators rather than the energies as in~\cite{VaiJonPot12}. Even then the difference is $\sim$0.38, which is well below $\delta(\nu_{18})\approx 3.35$ necessary to increase the size of the error bar for $n=18$ in Fig.~\ref{fig:C6_coeff_pp} to include the monotonic behavior. 

Below, we carry out a more direct sensitivity analysis on the robustness of the pronounced $n = 18$ feature seen in Fig.~\ref{fig:C6_coeff_pp} based on an analysis of the errors inherent in our radial matrix elements. Specifically, we investigate how much we can change the diatom matrix element $\langle 5s18p\,{}^1\!P_1,\,5s18p\,{}^1\!P_1|V_{\rm dd}|5s18d\,{}^1\!D_2,\,5s17d\,{}^1\!D_2\rangle$ by varying the model potential parameters, and how much this would affect the $\widetilde{C}_6$ coefficient we report at $n=18$ in Fig.~\ref{fig:C6_coeff_pp}. 

We consider what happens to the one electron matrix element $M_1=\langle 18p|V_{\rm d}|17d\rangle$ when we vary the model potential describing the $d$-states. We will do this by first defining the ratio of two matrix elements that come from two different model potentials used to describe the $17d$ state; $M_1/M_2 \sim (n_1^*/n_2^*)^2$ where $M_2$ is the new matrix element obtained using the modified potential, and $n_1^*$, $n_2^*$ are the two effective $n$ from the two different model potentials corresponding to the $17d$ state. Here we are not varying the model potential describing the $18p$ state, and we calculate $M_1$ using the model potential we use in the paper which gives $n_1^*=15.04$ for the $17d$ state. 

Defining $\delta \nu = |n_2^* - n_1^*|$ as the variation of the quantum defect from one model potential to the other, we find that $\delta\nu \sim \pm n_1^* ((M_2/M_1)^{1/2}-1)$ or $M_1[1 + (\delta\nu/n_1^*)]^2 \lesssim M_2 \lesssim M_1[1 - (\delta\nu/n_1^*)]^2$. We will try to achieve $\delta\nu\sim 3.35$ with the original $n_1^*\sim 15.04$ for the $17d$ state from the model potential in Sec.~\ref{subsec:mod-pot}. Note that we are only varying $n^*$ in the $^1D_2$ series and leave the $^1P_1$ series intact, effectively distributing the variation $\delta\nu\sim 3.35$ across the whole matrix element $\langle 5s18p\,{}^1\!P_1,\,5s18p\,{}^1\!P_1|V_{\rm dd}|5s18d\,{}^1\!D_2,\,5s17d\,{}^1\!D_2\rangle$. This makes $\delta\nu\sim 3.35$ into a parameter characteristic to the whole matrix element and not just the shift in the quantum defect in $17d$ state, and it is different than the actual shift in the quantum defect in any single one-electron  state. 

This approach gives $M_2\sim 1.49\,M_1$ for the upper bound, and we can vary one of the parameters in our model potential for the $^1D_2$ series so that the one electron matrix element is $M_2 = \langle 18p|V_{\rm d}|17d\rangle \sim 1.49\, M_1$ where $M_1$ is the matrix element obtained using the original model potential. Note that this change in the potential describing the $^1D_2$ series also affects the $18d$ state in the diatomic matrix element, changing its effective $n$ as well. The actual shift in the quantum defects due to this variation of the potential parameter for the $17d$ and $18d$ states are about $\sim$0.2, but shifts as large as $\sim$0.8 do not change our conclusions either as we will explain below. The reason these shifts are smaller than $\delta\nu\sim 3.35$ is because by shifting only the $d$-states and leaving the $p$-states intact varies the radial matrix elements faster due to faster reduction in the overlap of these states, compared to, for example, when both the $d$ and $p$ states are shifted together. 

When we calculate the $\widetilde{C}_6$ coefficients using the varied model potential, we find that the peak seen in Fig. 4 at $n=18$ is reduced by $\sim$55\% which places it slightly outside of the upper edge of the error bar coming from experimental energy uncertainties alone. In other words, by varying the one-electron matrix elements in the right hand side of $\langle 5s18p\,{}^1\!P_1,\,5s18p\,{}^1\!P_1|V_{\rm dd}|5s18d\,{}^1\!D_2,\,5s17d\,{}^1\!D_2\rangle$, we can only reduce the size of the $n=18$ peak in Fig. 4 by about half, leaving it still significantly larger than the $n^{11}$ scaling (blue diamonds in Fig. 4). However, the error bars are also reduced by about 50\% still leaving the $n^{11}$ prediction well outside. 

As we point out in Section III(C), the fractional differences between the experimental and the calculated energies, $\delta E/E^{\rm exp}_{np}$ and $\delta E/E^{\rm exp}_{nd}$, are at the $\sim\!10^{-2}$ level. 
This energy difference between the experimental and our calculated energies gives us a difference between the experimental and calculated quantum defects of about $\sim$0.2, about the same size in our discussion above. This means although the size of the peak can become smaller by about half due to the error in the radial matrix element, we find that the error bars also become smaller by about the same factor, making the peak a robust feature. 

Finally, varying the parameter $A$ in Eq.~\eqref{eq:modelPot} further, we can achieve a $\sim$0.8 shift in the actual quantum defects of the $17d$ and $18d$ states, where the $p$ states remain unshifted. In these cases, we find that the size of the peak at $n=18$ is further reduced to $|\widetilde{C}_6|\sim 4$ with significantly reduced error bars, still leaving the blue diamonds in Fig. 4 outside. 
Although the radial matrix element can be reduced by decreasing the overlap between the $d$ and the $p$ states leading to a reduction in $|\widetilde{C}_6|$ in this way, this reduction is accompanied by a reduction in the size of the corresponding error bar, ultimately leaving the $n^{11}$ result outside.

\bibliographystyle{prl2012}
\bibliography{LongRange_2026_2}

\end{document}